\documentclass[%
superscriptaddress,
 reprint,
preprint,
 amsmath,onecolumn,amssymb,
 aps,
12pt,letter]{revtex4-2}

\usepackage[english]{babel}
\usepackage{graphicx}
\usepackage{dcolumn}
\usepackage{bm}
\usepackage{float}
\usepackage{gensymb}
\usepackage[utf8]{inputenc}
\usepackage[shortlabels]{enumitem}
\usepackage{xcolor} 
\usepackage{amsmath}
\usepackage[export]{adjustbox}
\usepackage[colorlinks=true, allcolors=blue]{hyperref}
\usepackage{titlesec}
\titleformat*{\section}{\centering\normalfont\bfseries}
\titleformat*{\subsection}{\centering\normalfont\bfseries}
\titleformat*{\subsubsection}{\centering\normalfont}

\begin{document}

\title{An Assessment of Thermally Driven Local Structural Phase Changes in 1$T$$'$- MoTe$_2$}

\author{S. Khadka}
\affiliation{Department of Physics and Texas Center for Superconductivity, University of Houston, Houston, Texas 77204, United States}
\author{L.C. Gallington}
\affiliation{X-ray Science Division, Advanced Photon Source, Argonne National Laboratory, Argonne, Illinois 60439-4858, United States}
\author{B. Freelon}
\affiliation{Department of Physics and Texas Center for Superconductivity, University of Houston, Houston, Texas 77204, United States}


\begin{abstract}
The role of layer disorder is important in establishing the topological phases of MoTe${_2}$. A rich tapestry of atomic ordering influences the structural phase transitions (SPTs), but there is little understanding of the mechanistic details of the phase transition. An atomistic level study was conducted to investigate the local structure of the 1$T$$'$ and $T_d$ phases of MoTe${_2}$ by using the Pair Distribution Function (PDF) technique. While the average structure exhibits an SPT and co-existence of phases as a function of temperature, the local structure shows the suppression of SPT at short-range ordering, and the sample remains in the monoclinic (1$T$$'$) phase at all measured temperatures. A sharp PDF peak observed at short distances indicated a strong atom-atom correlation between the Mo and Te atoms within the Mo-octahedra. In addition, a large box modeling of the PDF data indicated a preferential motion of Te atoms towards $c$-axis at all temperatures. The structural defects, such as stacking faults, likely result in the co-existence of phases in the average structure and suppress the local SPT of MoTe$_2$. These results are stepping stones in understanding the long-debated origins of structural, vibrational, and electronic properties of MoTe$_2$ and similar transition metal dichalcogenides. 
\end{abstract}

\maketitle

\section{INTRODUCTION}
Transition metal dichalcogenides (TMDs) exhibit intriguing physical properties governed by planes of transition metal atoms arranged in graphene-like layers that interact through weak van der Waals coupling. \cite{jariwala_emerging_2014} Among the TMDs, MoTe$_{2}$ exhibits exotic properties such as large magnetoresistance, topological phases, and hosting Weyl fermions. It has also been extensively studied for tunability and enhancement of these physical properties. \cite{tamai_fermi_2016} Structural modulations can be achieved by changing the temperature, pressure, strain, or light, and it has been argued that MoTe$_{2}$ can serve as a topological switch based on these stimuli. \cite{dahal_tunable_2020, heikes_mechanical_2018, zhang_light-induced_2019} The high tunability has found effective applications in various fields, including field effect transistors (FETs), gas sensors, and water-splitting catalysts. \cite{fathipour_exfoliated_2014,iqbal_tailoring_2019,liu_atomic_2021}

\begin{figure}[!hb]
    \includegraphics{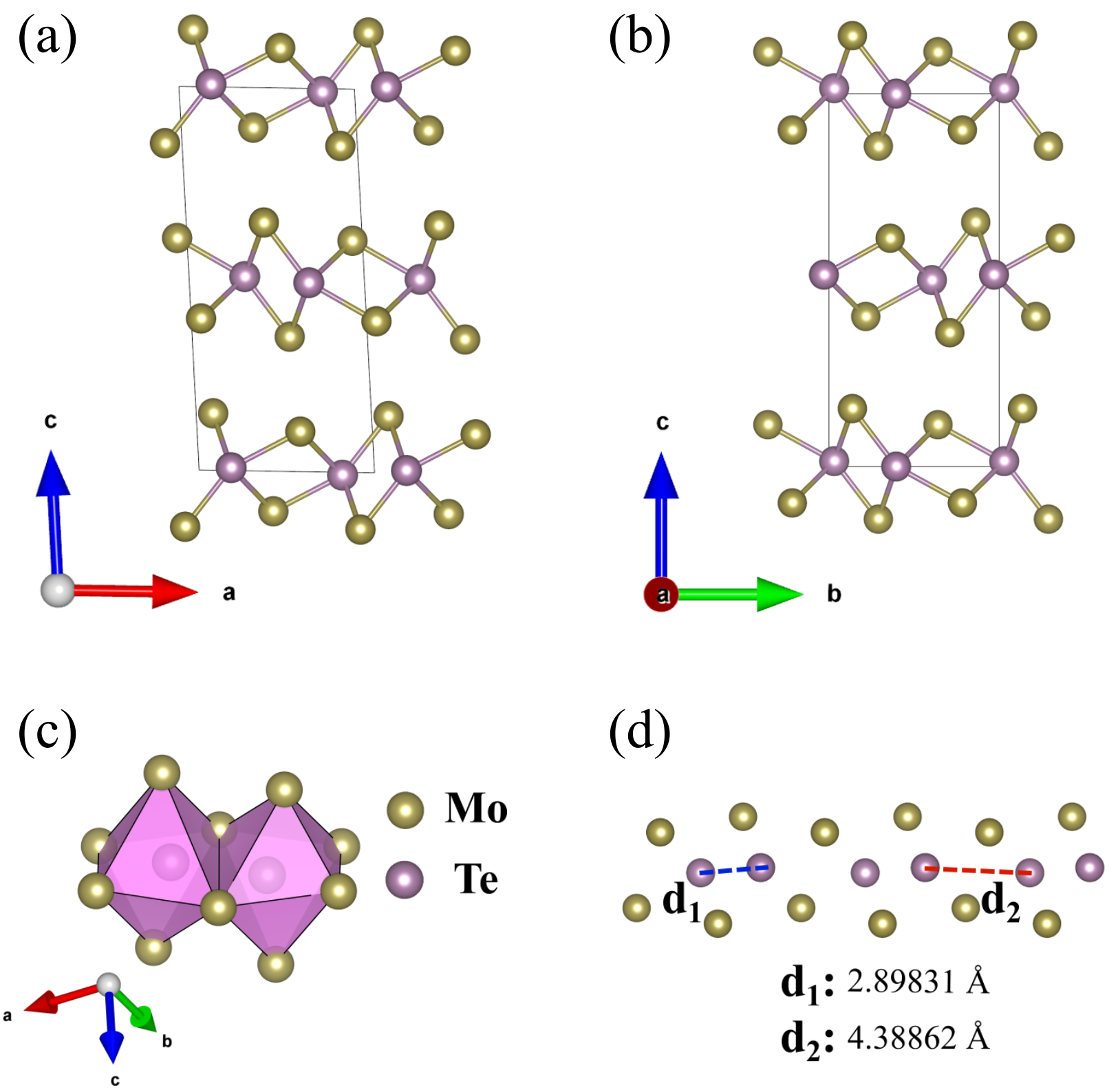}
    \caption{(a) The crystal structure for 1$T$$'$ (S.G. P2$_1$/$m$) MoTe$_2$ projected along [010] direction and (b) $T_d$ (S.G. P$mn$2$_1$) MoTe$_2$ projected along [100] direction. Te atoms are indicated by golden color. (c) Edge sharing Mo octahedra in the topological phases of MoTe$_2$ (d) A single layer in topological MoTe$_2$ that shows the modulation of interatomic Mo-Mo distances and out-of-plane distortion along the $c$-axis. Modulating the interatomic distances leads to many interesting emergent properties in the topological phases of MoTe$_2$.}
    \label{figure1_layeringunitcell}
\end{figure}

MoTe$_{2}$ crystallizes as three stable structures; hexagonal 2$H$, monoclinic and centrosymmetric 1$T$$'$, and the orthorhombic and non-centrosymmetric $T_d$ phases as shown in Fig. \ref{figure1_layeringunitcell}. \cite{qi_superconductivity_2016, kim_origins_2017} The 1$T$$'$ phase (S.G. P2$_{1}/m$) is stable at room temperature and undergoes a first-order SPT to the $T_d$ (S.G. P$mn$2$_{1}$) at temperatures below 250 K. \cite{yang_structural_2017} The 1$T$$'$ (topological insulating) and the $T_d$ (Weyl semimetallic) structural phases, seen in Figs. \ref{figure1_layeringunitcell}(a) and \ref{figure1_layeringunitcell}(b) are composed of identical Te-Mo-Te layers stacked along the $c$-axis. \cite{sun_prediction_2015} Both structures consist of distorted edge-sharing Mo-octahedra, and modulation of interatomic distances is observed within each layer, as shown in Figs. \ref{figure1_layeringunitcell}(c) and \ref{figure1_layeringunitcell}(d). Structural transition between these topological phases is not sharp and exhibits a rather large co-existence region reported to range from roughly 233 K to 290 K. \cite{hughes_electrical_1978,yang_elastic_2017,clarke_low-temperature_1978,heikes_mechanical_2018} Tuning of these topological phases by using external parameters is possible because of the subtle structural difference ($\sim$4$^{\circ}$ tilt of the $c$-axis for the 1$T$$'$ phase) and a small energy difference ($\Delta E_{T_d-1T'}=0.40$ meV per unit cell) between them. \cite{dahal_tunable_2020,clarke_low-temperature_1978,karki_strain-induced_2020,sie_ultrafast_2019,kuiri_thickness_2020} 

Due to weak interlayer interactions, disorders such as stacking faults, crystal twinning, and domain fragmentation are often observed in MoTe$_2$. Imperfections such as buckling and polyhedral distortion can affect the 1$T$$'$ $\rightarrow$ $T_d$ phase transition. Although structural modulations in MoTe$_2$ are described using short-range spatial parameters such as the interatomic and interlayer distances, \cite{schneeloch_emergence_2019, kim_origins_2017, qi_traversing_2022,singh_engineering_2020} few studies have used local structure techniques. An extended X-ray absorption fine structure study (EXAFS) on 2$H$ MoTe$_{2}$ showed an anomalous behavior on cooling down to 70 K, attributed to the dominant effects of various vibrational acoustic modes. \cite{caramazza_temperature_2016} A local structure study by Ref. \cite{petkov_local_2021} on 1$T$$'$ MoTe$_{2}$ using PDF showed that the local structure undergoes a second order phase transition (1$T$$'$ $\rightarrow$ $T_d$) with the monoclinic phase persisting down to 150K, the lowest temperature of that study. Local structure studies using the PDF technique have also been applied to other quantum materials to uncover the onset of superconductivity, charge density waves, and static and dynamic disorders. \cite{philip_out--plane_2021,smith_crystal_2008,subires_order-disorder_2023,malliakas_square_2005,kim_local_2006}

This work describes total X-ray scattering methods used to study the average and local structures of MoTe$_{2}$ in the 1$T$$'$ and $T_d$ phases over a temperature range of 95 K to 300 K. On average, MoTe$_2$ was found to stabilize in the 1$T$$'$ phase at high temperatures and undergo an SPT to the $T_d$ phase at low temperatures consistent with previous results. \cite{heikes_mechanical_2018, qi_superconductivity_2016, schneeloch_emergence_2019} Signatures of the orthorhombic phase were observed even at room temperature, possibly due to the effects of powder sample preparation processes. The persistence of the monoclinic phase well down to low temperatures (95 K) in average structure was confirmed by peak splitting. The thermal evolution of the local structure was analyzed for both cooling and warming cycles. Contributions from strong atom-atom correlation and atomic displacement parameters (ADPs) were studied in detail by analyzing PDF data down to temperatures that are lower than previous reports. The high momentum transfer and lower temperature of the current PDF experimental set-up allowed the observation of peak splitting that was not present in previously published data. A suppression in local SPT was also observed, likely due to the effects of the local structural defects, such as stacking faults and strong atom-atom correlations.

\section{EXPERIMENTAL}
High-quality 1$T$$'$ MoTe$_2$ single crystals were obtained from HQ graphene. \cite{noauthor_mote2_nodate} The single crystals were ground into a fine powder with an average crystallite size of 10 $\mu$m using mortar and pestle. Synchrotron X-ray diffraction (XRD) and X-ray PDF experiments were performed at the 11-ID-B PDF beamline of the Advanced Photon Source (APS) using 86.703 keV (0.143 \AA) x-rays. Data were acquired out to a $Q_{max}$ of 35 \AA$^{-1}$  to obtain PDFs with high real space resolution, thereby facilitating the observation of more local structural features. A CeO$_{2}$ standard was used to calibrate the sample-to-detector distance and the tilt of the image plate relative to the beam path. The distance from the sample to the detector was 1000 mm and 180 mm for the synchrotron XRD and the PDF experiment, respectively. Scattering measurements for the empty Kapton capillary were performed under the same experimental conditions to obtain the instrumental background.

Scattered intensities were recorded using a 2D amorphous Si detector in a Debye-Scherrer geometry. XRD data were collected at seven temperatures of interest between 300 K and 90 K, while PDF data were collected over the same temperature range with a difference of 5 K between each consecutive temperature. To investigate the local structure behavior in detail, we recorded PDF data at temperature intervals of 2 K close to the nominal bulk phase transition temperature ($\sim$250 K). XRD and PDF data were collected during the cooling and warming of the sample. The temperature was controlled with liquid nitrogen using the Oxford Cryostream 800+ system. 

Intensity data versus 2$\theta$ and $Q$ were obtained by converting the integrated image using the GSAS-II software, and PDF data were generated using PDFgetX3 routines.\cite{toby_gsas-ii_2013,farrow_pdffit2_2007} Data corrections, including background subtraction, sample self-absorption, multiple scattering, X-ray polarization, Compton scattering, and Laue diffuse scattering, were implemented to obtain the normalized total scattering structure function $S(Q)$. \cite{juhas_pdfgetx3_2013} Ni standard data were processed and refined to obtain the instrumental damping factor for the PDF data. \cite{pourpoint_new_2012}

\section{RESULTS}\label{resultsLabel}
\subsection{Average Structure}
\begin{figure}[t]
    \includegraphics[scale=0.75]{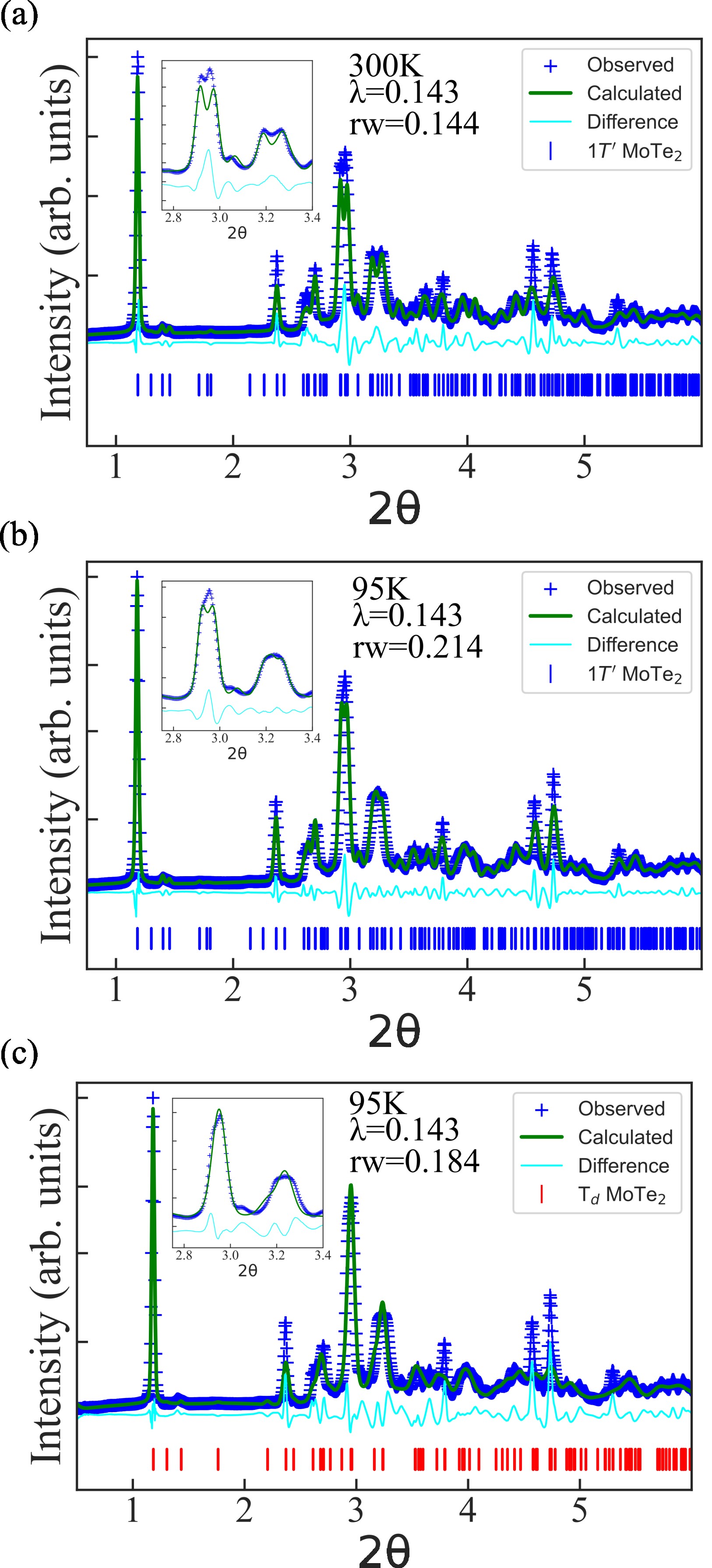}
    \caption{(a) and (b) Results from RR using the monoclinic (P2$_1$/m) phase for XRD data at 300 K and 95 K, respectively. The merging of the peaks at 2$\theta$=2.954$^{\circ}$ and 2$\theta$ = 3.236$^{\circ}$ at 95 K is not reproduced well by this phase. (c) Rietveld refinements using the orthorhombic (P$mn$2$_1$) phase. The merging of the peaks at 2.954$^{\circ}$ and 3.236$^{\circ}$ are well reproduced by this phase.}
    \label{refinementXRD_00}
\end{figure}

The X-ray powder diffraction intensity profile of the MoTe$_2$ sample at 300 K, together with the results of the Rietveld Refinement (RR), is shown in Fig. \ref{refinementXRD_00}(a). Rietveld refinement allows the precise determination of atomic positions and other instrumental and structural parameters to resolve the crystal structure of a material by fitting a theoretical model to experimental data. \cite{rietveld_rietveld_2014, runcevski_rietveld_2021} 

Using the GSAS II package, RR of the powder pattern (Fig. \ref{refinementXRD_00}(a)) collected at 300 K revealed the structural symmetry to be monoclinic (S.G. $P2_1/m$) commonly known as the 1$T$$'$ phase. At higher temperatures, the peak at 2$\theta$ = 2.954$^{\circ}$ splits and can be indexed as ($\bar{1}$12)$_\textrm{M}$ and (112)$_\textrm{M}$ where the subscript $\textrm{M}$ indicates the monoclinic symmetry. Another splitting at 3.236$^{\circ}$ is also observed and can be indexed as ($\bar{1}$13)$_\textrm{M}$ and (113)$_\textrm{M}$, confirming the 1$T$$'$-type unit cell for the sample at high temperatures. \cite{petkov_local_2021} These peaks merge into a single peak as the temperature is lowered. The Rietveld refinements performed using the monoclinic and the orthorhombic phases for XRD data at 95 K are shown in Fig. \ref{refinementXRD_00}(b) and \ref{refinementXRD_00}(c) respectively. The merged peaks can be indexed only by the orthorhombic (P$mn$2$_1$) unit cell and can be indexed as (112)$_\textrm{O}$ and (113)$_\textrm{O}$. However, the orthorhombic phase did not yield better  $R_{wp}$ and $\chi^2$ values than the monoclinic phase, indicating that the monoclinic phase still persists even well below the reported SPT temperature (250 K). Table~\ref{tab:SinglePhaseRR} presents the results from single-phase RR.

\begin{figure}[t]
    \includegraphics[scale=0.75]{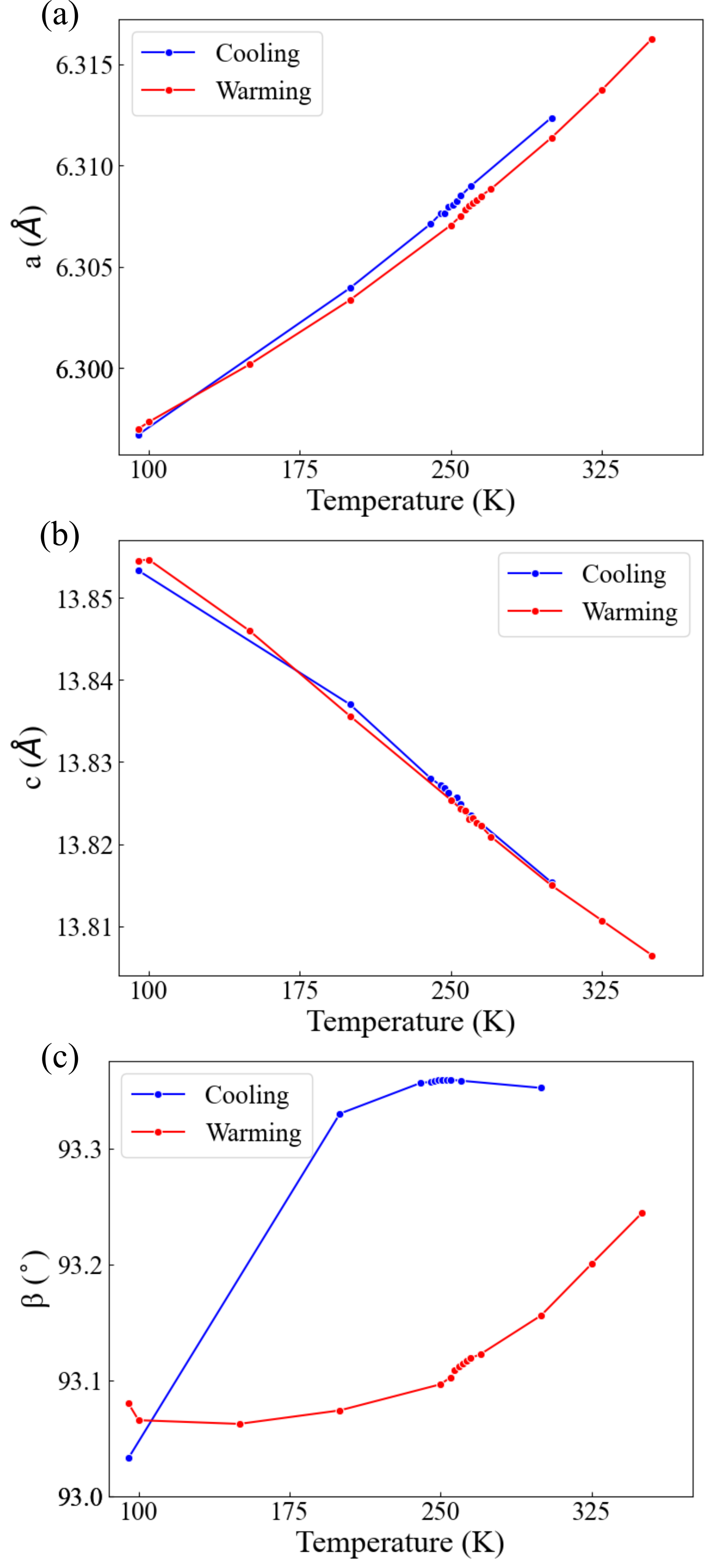}
    \caption{(a) - (c) Temperature dependence of lattice parameters $a$, $c$, and $\beta$, respectively, obtained after RR using the monoclinic unit cell. Lattice parameters $a$ and $c$ do not show significant discontinuity or thermal hysteresis. $\beta$ shows a discontinuity and thermal hysteresis reminiscent of a 1$T$$'$ $\rightarrow$ $T_d$ phase transition.}
    \label{latticeParams}
\end{figure}

The lattice parameters obtained from RR using the monoclinic phase are shown in Figs. \ref{latticeParams}(a) - \ref{latticeParams}(c). The lattice parameters $a$, $b$, and $c$ did not exhibit significant discontinuity. However, the $\beta$ angle showed a discontinuity around 200 K. A thermal hysteresis was also observed for the $\beta$ angle. Thermal hysteresis in vdW materials has been attributed to stacking faults \cite{tao_appearance_2019, hovden_atomic_2016, yan_composition_2017}. The transition barrier between different stacking sequences of 1$T$$'$, $T_d$, or random stacking sequences in MoTe$_2$ could be responsible for the observation of the thermal hysteresis. \cite{lee_origin_2019,hart_emergence_2022} 

\begin{table}[!h]
\caption{\label{tab:SinglePhaseRR}Lattice parameters ($a$, $b$, $c$, $\beta$, and unit cell volume (Vol.)) and quality of fits ($R_{wp}$ and goodness of fit (G.O.F.)) for RR using a single phase.}
\setlength\tabcolsep{1pt}
\begin{ruledtabular}
\begin{tabular}{lcccccccr}
Temp (K) & S.G. & \textit{a}(\AA) & \textit{b}(\AA) & \textit{c}(\AA) & $\beta$($^{\circ}$) & Vol.(\AA$^3$) & $\mathrm{R}_{wp}$ & G.O.F\\
\colrule
300 & P2$_1$/$m$ & 6.312(12) & 3.4680(5) & 13.8154(27) & 93.352(14) & 301.9(6) & 14.41 & 8.44 \\
95 & P2$_1$/$m$ & 6.2967(15) & 3.4591(6) & 13.8533(3) & 93.033(17) & 301.3(8) & 18.36 & 11.64 \\
95 & P$mn$2$_1$ & 3.4591(10) & 6.2647(3) & 13.8087(28) & 90 & 299.24(15) & 21.37 & 13.41 \\
\end{tabular}
\end{ruledtabular}
\end{table}

\begin{figure}[t]
    \includegraphics[scale=0.75]{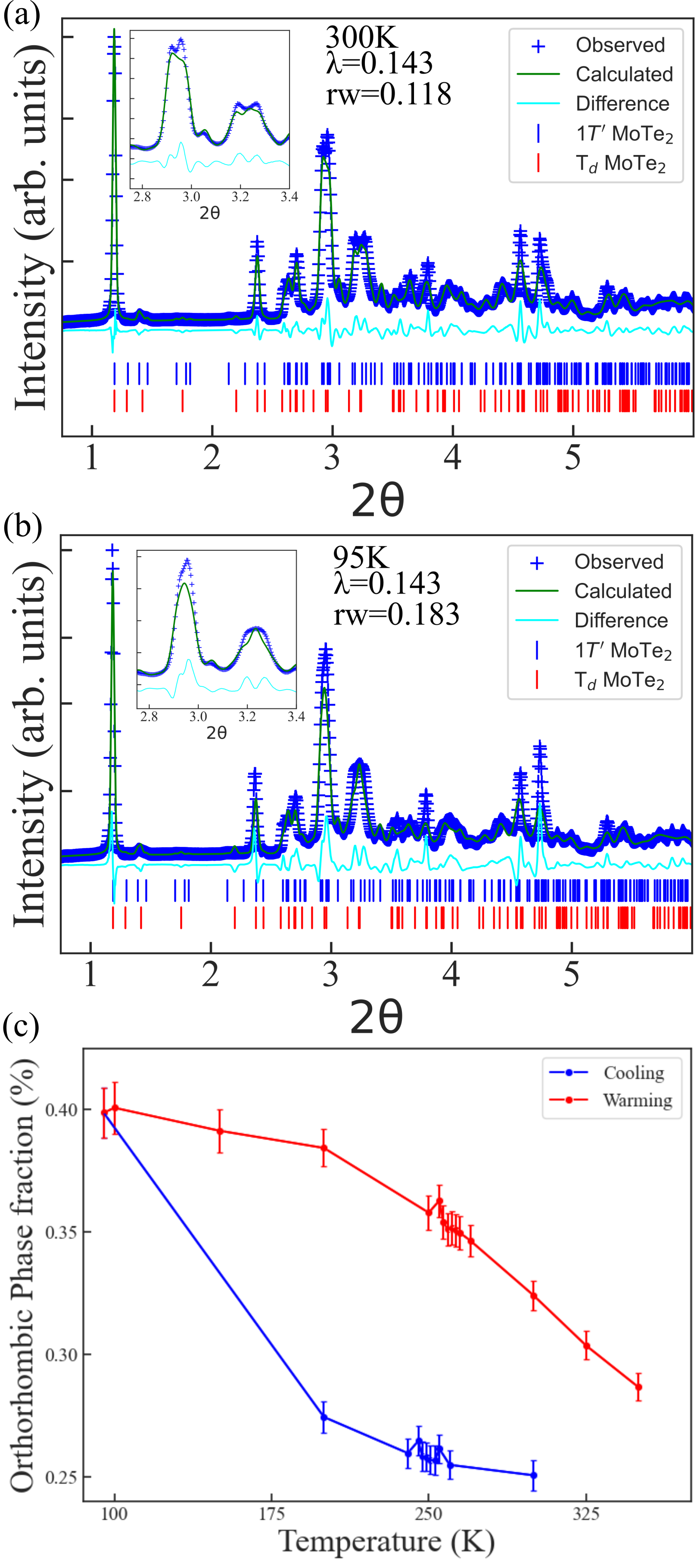}
    \caption{(a) Multiphase RR using monoclinic and orthorhombic phases shows reasonable agreement with the experimental data, indicated by the comparable $R_{w}$ values to those obtained by single-phase RR. (b) Multiphase RR for 95K. (c) Phase fraction of the orthorhombic phase in the sample at various temperatures during cooling and warming the sample. The phase fraction for the orthorhombic phase is $\sim$0.25, even at room temperature.}
    \label{multiphaseRef_Results}    
\end{figure}

The possibility of multiple phases in the sample \cite{clarke_low-temperature_1978} was investigated by using multiphase RR, as shown in Fig. \ref{multiphaseRef_Results} and its results are presented in Table ~\ref{tab:multiPhaseRR}. During multiphase refinement, the relative amounts of each phase present in the sample, phase fraction, \cite{toby_gsas-ii_2013} were refined to systematically investigate their contributions to the observed scattering pattern. Refinements of background, instrument parameters, and sample parameters were performed similarly to single-phase refinements. The multiphase RR for XRD data collected 300 K is presented in Fig \ref{multiphaseRef_Results}(a). Sequential refinements showed that an orthorhombic phase is present even at high temperatures and increases with decreasing temperatures. \cite{dahal_tunable_2020, heikes_mechanical_2018, kim_origins_2017} The layers in MoTe$_{2}$ can easily slide relative to each other because of weak in-plane coupling \cite{cheon_structural_2021} and can potentially form orthorhombic stacking order in between the 1$T$$'$ layers even at higher temperatures. \cite{schneeloch_emergence_2019, qi_traversing_2022} Defects and variability in powder sample preparation processes can also result in orthorhombic stacking even at higher temperatures. \cite{schneeloch_antiferromagnetic-ferromagnetic_2023} Phase fraction of the orthorhombic phase at different temperatures is shown in Fig. \ref{multiphaseRef_Results}(b) which increases from $\sim$0.25 at 300 K to $\sim$0.40 at 95 K. This indicated that more fraction of the sample has transitioned to the orthorhombic phase. Upon warming, the phase fraction does not return to the initial value of $\sim$0.25 even at 350 K, which may be due to thermal hysteresis in the sample. \cite{clarke_low-temperature_1978}

\begin{table}[!h]
\caption{\label{tab:multiPhaseRR}Quality of fits for multiphase RR using both 1$T$$'$ and $T_d$ phases}
\begin{ruledtabular}
\begin{tabular}{lcr}
Temp (K) & $R_{wp}$ &GOF\\
\colrule
300 & 11.76 & 6.93 \\
95 & 18.33 & 11.68 \\
\end{tabular}
\end{ruledtabular}
\end{table}

\subsection{Local Structure}
\subsubsection{Small Box Modeling}
Recent studies have emphasized the importance of local structure in cases of van der Waal low-dimensional materials and have shown how real-space structure analysis can provide intriguing insights into the average structure. \cite{egami_local_2004, billinge_rise_2019, dove_review_2022} Small box modeling using PDFGui software was employed for a Rietveld-like analysis of the origins of the temperature dependence of the PDF profiles. PDFGui deploys a Rietveld-like refinement technique to PDF data. $S(Q)$ data were reduced to observed $G(r)$ data using PDFgetX3 software. Suitable values of $Q_{min}$, $Q_{max}$, and background coefficients were corrected to obtain the best possible $G(r)$ data that was void of any artifacts. $G(r)$ was refined using PDFGui by systematically adjusting scale factors, lattice parameters, atomic positions, and ADPs. Other parameters, such as atom-atom correlation factors and instrument dampening parameters, were refined if deemed to improve the fit quality significantly. The quality of refined (goodness of fit) is calculated by normalizing the difference between the experimental and calculated values and is indicated by the parameter $R_{wp}$. \cite{farrow_pdffit2_2007}
\begin{figure}[b]
    \includegraphics[scale=0.85]{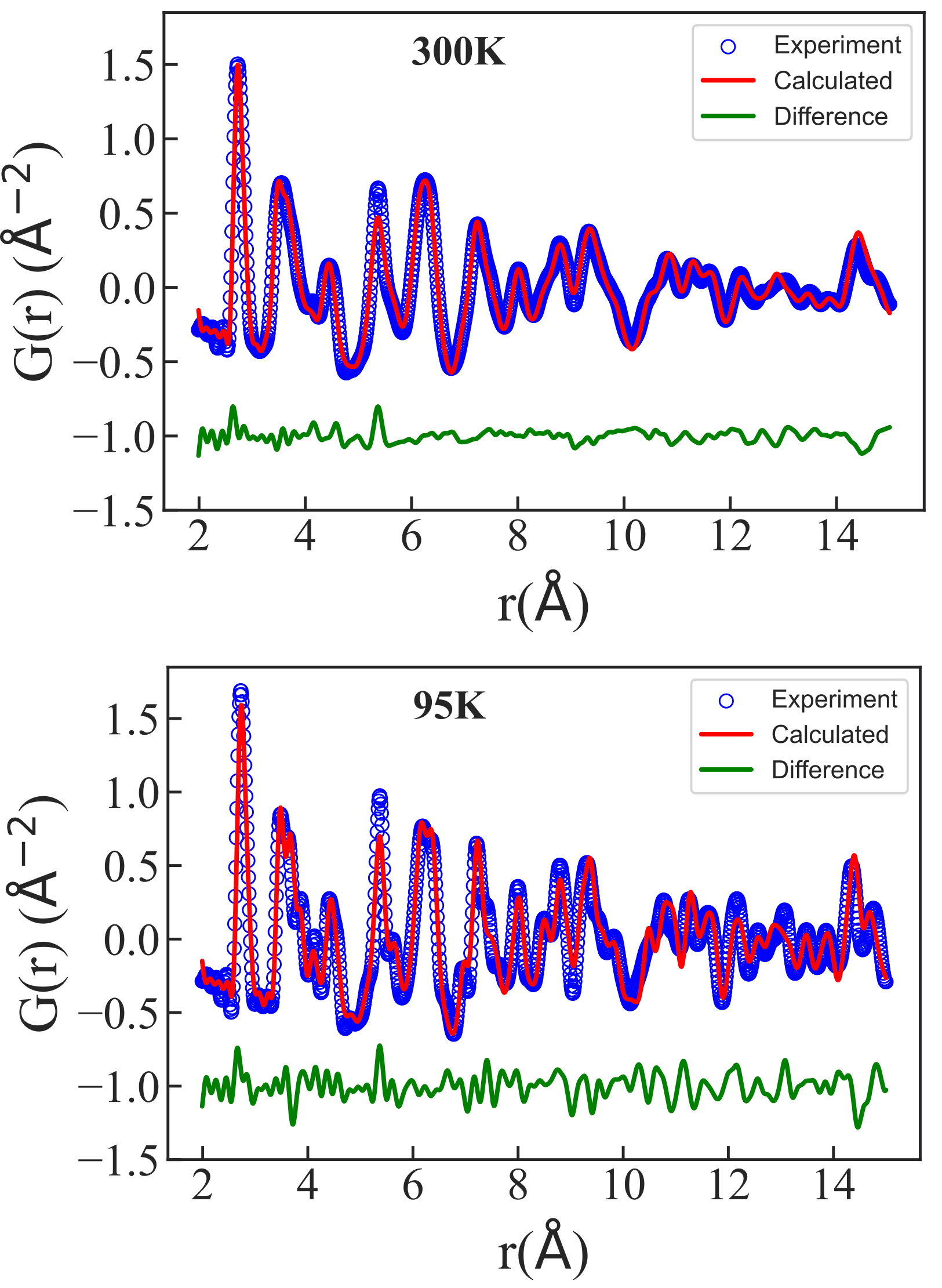}
    \caption{(a)PDF Fits at 300 K using small box modeling. Blue open circles represent experimental data points, and the red curve represents the calculated pattern, while the green curve represents the residual. Residual (difference between experiment and calculated $G(r)$) is shifted vertically for clarity. Monoclinic (P2$_1$/$m$) unit cell was used for refinements at both temperatures. (b) PDF Fits at 95 K using small box modeling. The fit worsens for low-temperature refinements, but the $R_{wp}$ value is still acceptable with the monoclinic unit cell.}
    \label{PDFGui_Fitting_cooling}
\end{figure}

Fig. \ref{PDFGui_Fitting_cooling}(a) shows the PDFGui fitting for MoTe$_2$ at 300 K using the monoclinic 1$T$$'$ structure. After refining the scale parameters, lattice parameters, and atomic positions, an $R_{wp}$ value as low as 14\% was obtained. The PDFGui fitting presented in Fig. \ref{PDFGui_Fitting_cooling}(b) for the low temperature (95 K) is reasonable using the monoclinic unit cell. This finding reveals that despite the SPT observed in the average structure analysis, the local structure does not undergo observable structural changes besides the systematic changes in the refinement parameters. Since the atomic arrangement for both topological phases consists of corrugated Te-Mo-Te layers stacked along the crystallographic $c$-axis with a very small energy difference, the structural distinctions between these phases at short distances (within a single unit cell, 0 - 15 \AA) are negligible. Using the orthorhombic phase resulted in no observable improvement in the fit quality, as seen by comparing $R_{wp}$ value (See section of IV of Supplemental Material (SM) at \cite{khadka_assessment_nodate}). 

\begin{figure}[!ht]
     \includegraphics[scale=0.65]{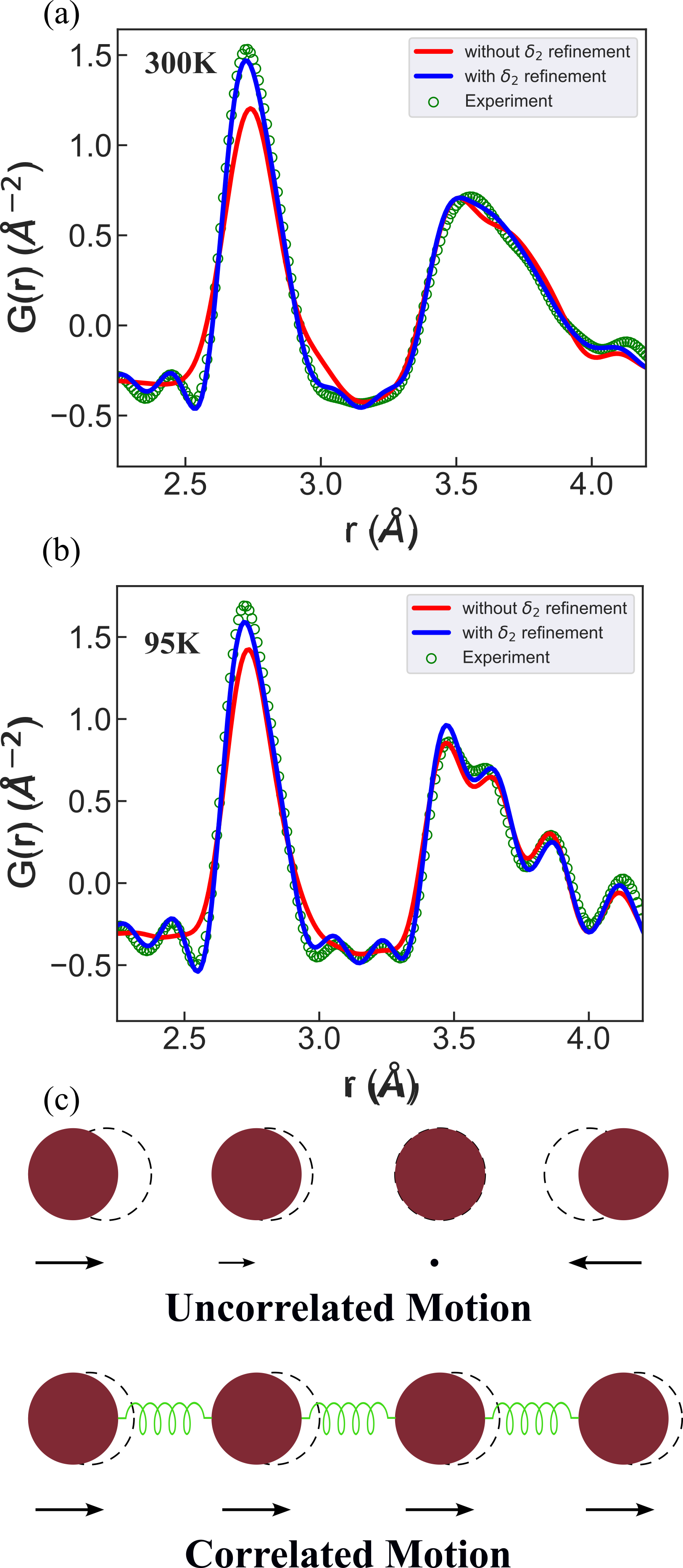}
     \caption{(a) Small box refinements for MoTe$_2$ PDF data obtained at room temperature (300 K) with and without $\delta_2$ refinements. (b) Refinements for PDF data at low temperatures (95 K). In both figures, the red curve indicates refinement without using the $\delta _2$ parameter, the blue curve shows refinement using the $\delta _2$ parameter, and the green circles show the experimental values. For both temperatures, refinement of $\delta_2$ significantly improves the quality of fits. The sharp first peak at $r$ = 2.72 \AA \ is fit only when this parameter is refined, indicating a strong atom-atom correlation between Mo and Te atoms. (c) Schematic diagram showing a snapshot of atomic positions due to uncorrelated and correlated motion. The arrow indicates the motion of atoms, and their new position is shown as dashed circles. The force of interaction between strongly correlated atoms can be considered a spring force. \cite{jeong_lattice_2003}}
     \label{deltaContribution}
\end{figure}

For both temperatures, a sharp first peak corresponding to the Mo-Te bond within the Mo octahedra (see Fig. \ref{figure1_layeringunitcell}(c)) was observed. The movement of the Mo and Te atoms can be strongly correlated because of the interatomic forces that depend on the atomic pair distances illustrated in Fig. \ref{deltaContribution}(c). This force is well-described only for the first few nearest neighbors, thereby producing a sharp peak at low $r$ values. \cite{dove_review_2022,jeong_lattice_2003} The correction to the peak widths ($\sigma_{ij}$) is performed by introducing $\delta _1$ and $\delta _2$ parameters in the PDFGui refinements. The final peak width, including corrections, is given by 
\begin{equation}
    \sigma_{ij}=\sigma_{ij}'\sqrt{1-\frac{\delta_1}{r_{ij}}-\frac{\delta_2}{r_{ij}^2}+Q_{broad}^2r_{ij}^2}
\end{equation}
where $\sigma_{ij}'$ is the peak width without correction and $Q_{broad}$ is the peak broadening as a result of the $Q$ resolution of the instrument. \cite{farrow_pdffit2_2007} The first peak was well fitted only when the correlation parameter $\delta_{2}$ was refined, as shown in Fig. \ref{deltaContribution}, indicating the presence of strong atom-atom correlation between Mo and Te atoms within the octahedra. 

\begin{figure}[t]
    \includegraphics{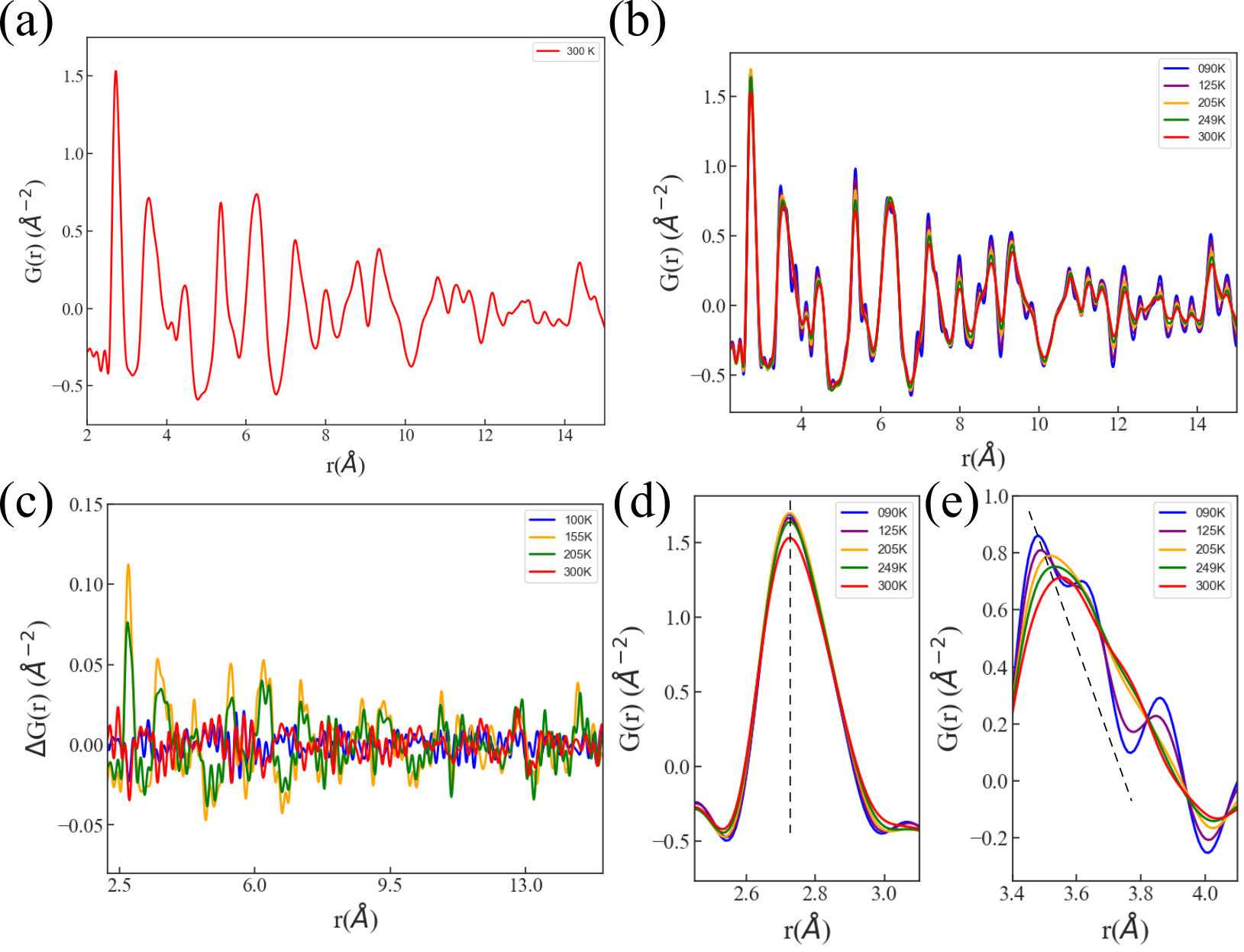}
    \caption{(a) $G(r)$ at 300 K up to 15 \AA. Sharp peaks are observed, confirming the crystallinity of the sample. (b) $G(r)$ up to $r$ = 15 \AA \ at different temperatures during cooling cycle. (c) The intensity difference between PDFs obtained at the same temperature during the cooling and warming of the sample shows hysteresis across the temperature range of the study, (d) - (e) Magnified view of the first and second peak of PDF during the cooling cycle shows temperature evolution of the Mo-Te bond ($r$ = 2.72 \AA) and the Te-Te ($r$ = 3.54 \AA) correlations, respectively.}
    \label{baselinePDF}
\end{figure}

Fig. \ref{baselinePDF}(a) shows the MoTe$_2$ PDF data collected at 300 K in the range $r$ = 0 - 15 \AA, and well-defined peaks are observed well up to $r$ = 40 \AA \, which confirms the crystalline character of the material. The peaks at larger distances ($r>$ 25 \AA) indicate the presence of long-range correlations in the material but are broadened compared to the peaks at shorter $r$. The first peak in Fig. \ref{baselinePDF}(a) at $r$ = 2.72 \AA \ indicates the coordination environment due to the Mo-Te bonds in MoTe$_2$ while the broad second peak at $r$ = 3.2 \AA \ develops due to contributions from both Mo-Mo and Te-Te bonds. The temperature evolution of the PDF peaks is shown in Fig. \ref{baselinePDF}(b). At lower temperatures, the splitting of peaks is observed ($r$ = 3.2 \AA). Significant hysteresis is evident from Fig. \ref{baselinePDF}(c), which shows differences in the PDF peaks at the same temperature during the cooling and warming of the sample.

The center of the first peak at $r$ = 2.72 \AA \ in the PDF corresponds to the intralayer Mo-Te distance and does not change significantly during thermal cycles. Minor changes in the peak height and width are observed as shown in Fig. \ref{baselinePDF}(d) and indicate an insignificant intralayer distortion in MoTe$_{2}$ during thermal cycles. However, the effects of temperature cycling on the shape of the second peak are clear, as shown in Fig. \ref{baselinePDF}(e). A peak splitting occurs upon cooling. The new peaks at 3.49 \AA \ and the other at 3.62 \AA \ correspond to the interlayer Mo-Mo and Te-Te distances, respectively. The peaks are distinctly separated at very low temperatures and indicate contributions from ADPs on the local structure ordering in MoTe$_{2}$. 

\begin{figure}[t]
     \includegraphics{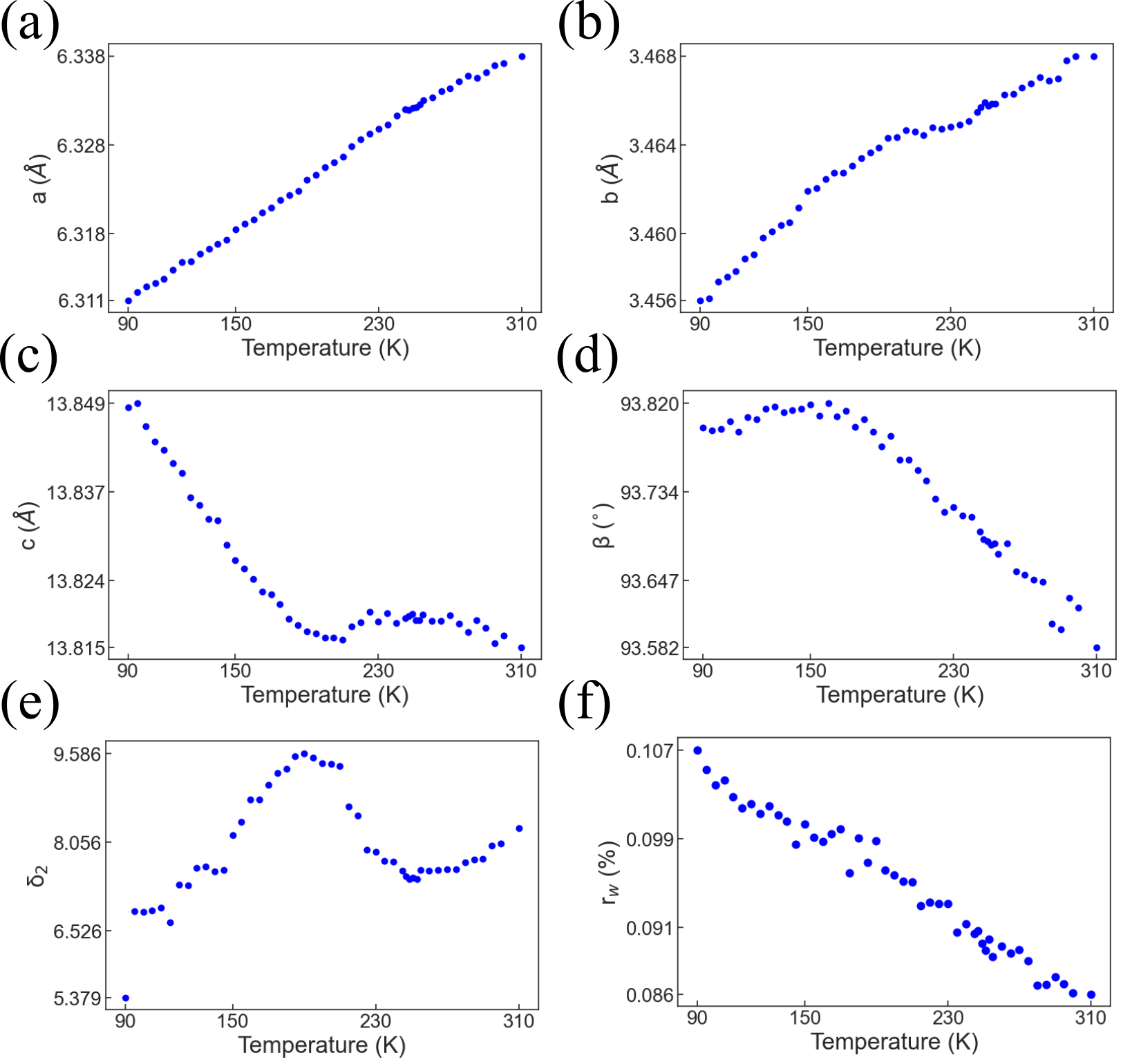}
     \caption{(a) - (c) Temperature dependence of the lattice parameters, $a$, $b$, $c$, derived from small-box refinement. $a$ and $b$ show positive thermal expansion while $c$ shows negative thermal expansion. The temperature dependence of (d) $\beta$ angle, e) atom-atom correlation ($\delta_2$) parameter, and (f) goodness of fit ($r_w$) for the cooling cycle from small box modeling.}
     \label{PDFGui_LatParams}
\end{figure}

Temperature series refinement was completed for cooling and warming cycles. The temperature dependence of the lattice parameters and the quality of fit ($R_{wp}$ are shown in Fig. \ref{PDFGui_LatParams}. The quality of fit gradually deteriorates when using PDF data at lower temperatures, and it is 10.5 \% at 95 K(see Fig. \ref{PDFGui_LatParams}(f)). Lattice parameters $a$, $b$, and $\beta$ exhibit a positive thermal expansion, while lattice parameter $c$ shows a negative thermal expansion, consistent with previous experimental observations. \cite{kim_thermomechanical_2021, heikes_mechanical_2018} The $c$-lattice and $\beta$ angle display a discontinuity near temperatures approaching 200 K, unlike the continuous variations observed in the $a$ and $b$ lattice parameters throughout both cooling and warming cycles.

\subsubsection{Large Box Modeling}
\begin{figure}[h]
     \includegraphics{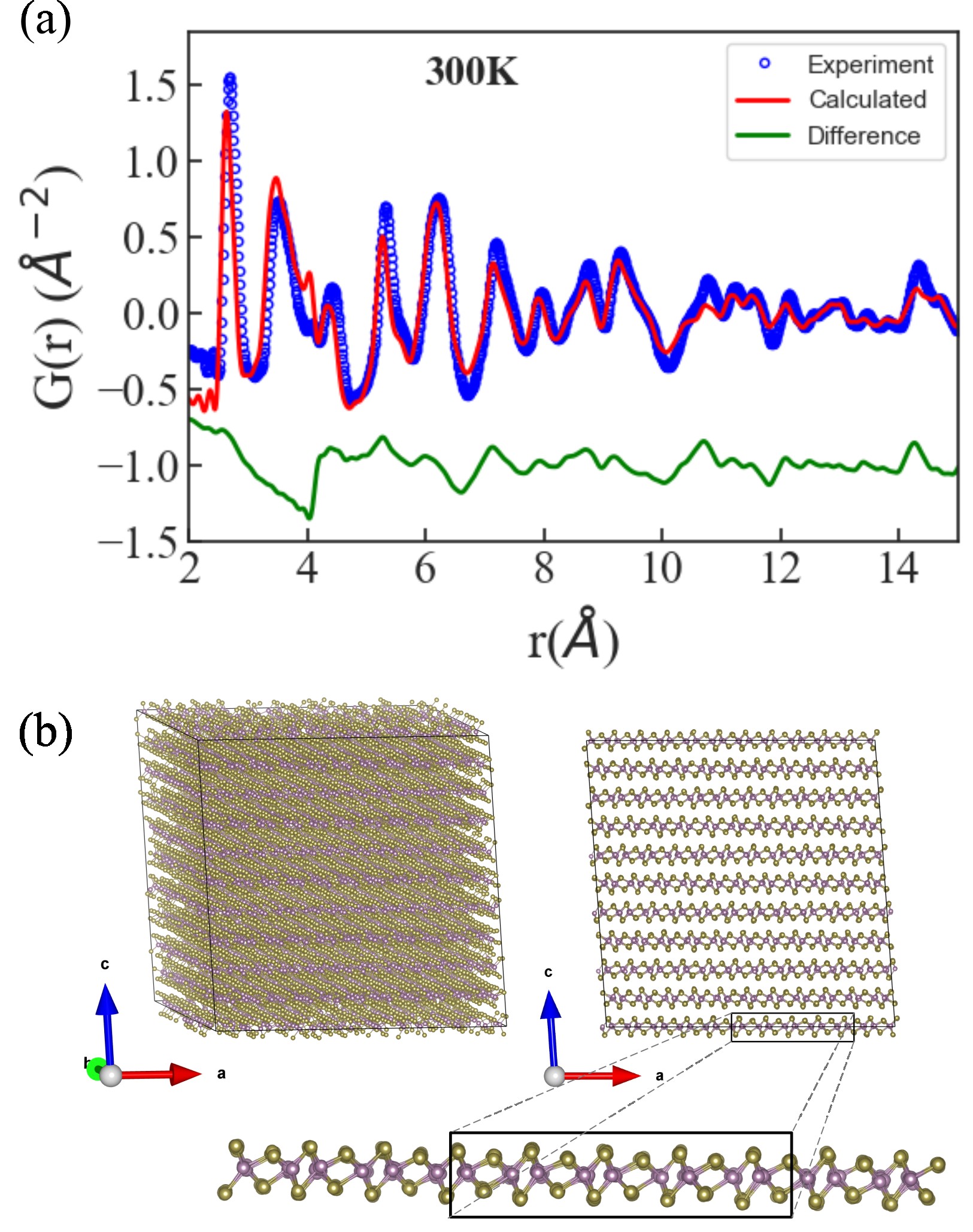}
     \caption{(a) Fit for 300K PDF data using a 10x20x5 monoclinic supercell. The supercell contained 13200 atoms (4400 Mo, 8800 Te atoms). Blue circles represent the experimental data, red circles show the simulated PDF from RMCProfile, and green circles represent their difference. The difference curve has been shifted vertically for clarity. (b) Snapshots of MoTe$_2$ supercell obtained after RMCProfile refinement was completed. Perspective view and view along $b$-axis. A magnified view of a portion of a layer (viewed along $b$-axis) indicates the presence of position disorder for each atom.}
     \label{RMCRefinedSupercell}
\end{figure}

Small box modeling of the PDF data indicated suppression of the SPT on MoTe$_2$ and indicated a strong Mo-Te correlation based on the sharp peak behavior at $r$ = 2.72 \AA. Due to this strong atom-atom correlation, near-neighbour Mo-Te pairs are coupled in their movements and move in phase, whereas far-neighbor atomic pairs move almost independently. The large box modeling approaches more accurately capture these effects since the refinement is performed by systematically displacing individual atoms in a large supercell without prioritizing overall periodicity. Therefore, large box modeling can provide useful insights into the temperature dependence of correlated motion and dynamic disorder in MoTe$_2$. \cite{dove_review_2022} RMCProfile was used to refine the MoTe$_2$ X-ray PDF data at different temperatures using a $20\times 10\times 5$ supercell such that each side of the simulation box $\sim$ 65 \AA\ and contained 12000 atoms (4000 Mo and 8000 Te atoms). An indicator of the quality during the modeling/ refinement is $\chi^2$ expressed as $\chi^2=\sum_j (G(r)_j^{\text{exp}} - G(r)_j^{\text{calc}} )/\sigma_j^2$, where the summation is over all data points (labeled by $j$). As seen in Fig. \ref{RMCRefinedSupercell}(a), the agreement between the experimentally obtained and simulated PDF using RMCProfile modeling is reasonable. The goodness of fit shown in the figure is $\chi^2=5.07$, which indicates a good fitting quality. In contrast to small-box modeling, large-box modeling does not consider long-range ordering, thereby enabling the introduction of disorders in atomic positions. This approach uses the Metropolis Monte Carlo algorithm to produce atomic configurations consistent with experimental data. Atoms in the configuration are selected randomly and then displaced by a random amount, and the closeness of the fit to the experimental data is used to determine if each move is accepted. \cite{tucker_rmcprofile_2007} These routines within a large box modeling facilitate the introduction of various stacking and layering models (see section VI in SM at \cite{khadka_assessment_nodate} for details). A simple distance constraint in our simulations limits the bond length between two atoms in MoTe$_2$. The minimum distance for each of the bonds and the maximum displacement of each atom at each step were also explicitly mentioned in the simulation parameters. Other constraints, such as polyhedral and bond valence constraints, did not significantly improve the fits.

The snapshot of the atomic configuration after large box modeling at 300K is shown in Fig. \ref{RMCRefinedSupercell}(b). Prominent positional disorders and no other disorders, such as vacancies and rotation of layers were observed. A magnified view of a section of a Te-Mo-Te layer indicates no significant distortion to the Mo polyhedra at all temperatures in this study. The Mo polyhedra in MoTe$_2$ may act as a Rigid Unit Mode (RUM) and can translate or rotate without any significant distortion. \cite{tan_rigid_2023} RUMs are usually associated with the negative thermal expansion (NTE) observations, such as the NTE of $c$-axis in MoTe$_2$. \cite{tao_role_2003, dove_which_2023} The RUMs may also suppress the SPT, as observed in this study.

The displacement disorder is more clearly visualized with a projection of the supercell into the original unit cell (see section V in SM \cite{khadka_assessment_nodate} for more details). This approach revealed the preferential displacement of Te atoms along the $c$-axis, which is more noticeable at higher temperatures due to changes in interlayer distances. Refinement with a $T_d$ supercell similarly revealed Te atom displacement along the stacking direction, with the tilt occurring along the short axis in both the 1$T$$'$ supercell and $T_d$ supercell cases, thereby confirming the relative sliding motion of Te-Mo-Te layers along the short axis. \cite{schneeloch_emergence_2019}

\section{DISCUSSION}
Despite older and renewed research on MoTe$_2$, much is yet to be learned about the nature of its SPTs. Data in Fig. \ref{refinementXRD_00} demonstrates the bulk 1$T$$'$ $\rightarrow$ $T_d$ SPT due to temperature changes. Rietveld refinements using the monoclinic phase yielded the expected negative thermal expansion of the $c$-axis and was previously attributed to the phonon softening in MoTe$_2$. \cite{kim_thermomechanical_2021} The splitting of peaks at the low-temperature Xray data also provided evidence of the appearance of an orthorhombic $T_d$ phase consistent with previous studies \cite{heikes_mechanical_2018, clarke_low-temperature_1978}. However, evidence of the co-existence of the 1$T$$'$ and $T_d$ phases was also observed. Multiphase RR using the monoclinic 1$T$$'$ and orthorhombic $T_d$ phases in Fig \ref{multiphaseRef_Results} indicated the $T_d$ phase fraction to be $\sim$0.25 at 300 K. Defects and variability in powder sample preparation processes, such as in Ref. \cite{schneeloch_antiferromagnetic-ferromagnetic_2023}, might explain the appearance of orthorhombic stacking at higher temperatures. In this scenario, MoTe$_2$ layers can easily slide relative to each other because of weak in-plane coupling \cite{cheon_structural_2021} and could form orthorhombic stacking order in between 1$T$$'$ layers. \cite{schneeloch_emergence_2019, qi_traversing_2022} While the presence of an orthorhombic phase at room temperature in bulk MoTe$_2$ has not been reported in previous studies, observation of different possible stacking combinations in similar van der Waals materials, such as CrI$_3$, has been reported to be dependent on the powder sample preparation process. \cite{schneeloch_antiferromagnetic-ferromagnetic_2023} Another possibility is the existence of an orthorhombic phase that is not $T_d$ MoTe$_2$ but a pseudo-orthrombic phase. Tao \textit{et. al. }has reported one such phase called the $T_d^*$ and attributes the formation of this phase to the random change in the stacking sequences. Rietveld refinements using this orthorhombic phase $T_d^*$ performed just as well as the orthorhombic $T_d$ phase (see section VI of SM at \cite{khadka_assessment_nodate} for more details). This indicates that phases with pseudo-orthorhombic stacking with a phase fraction of $\sim$25\% are present due to the sample preparation process and other disorders.

The monoclinic phase was also observed down to 95 K, indicating that 1$T$$'$ layering is pinned and does not undergo a phase transition at lower temperatures. This explains the observation of a high monoclinic phase percentage even at low temperatures. Pinning of the monoclinic phase due to defects present in the sample can cause the persistence of monoclinic stacking at temperatures well below the nominal SPT temperature of 250 K. In this scenario, the MoTe$_2$ lattice gradually adjusts to the strain produced due to a change in temperature by splitting into small domains. \cite{clarke_low-temperature_1978} Local defects can prevent the rearrangement of these domains, thereby pinning the sample in the monoclinic phase even at lower temperatures. Hole doping from sample exposure to the atmosphere can also stabilize the 1$T$$'$ monoclinic domains in the sample,even at lower temperatures. \cite{kim_origins_2017,paul_tailoring_2020,he_dimensionality-driven_2018, mandal_enhancement_2018, sakai_critical_2016} The continuous change in the lattice parameters (except $\beta$) could also be due to this persistence of the monoclinic phase. 

With this understanding of the average structure of our samples, we investigated the local structure, and it was possible to model the PDF data using small-box and large-box modeling approaches. The PDF data showed structural features consistent with recently published results in Ref. \cite{petkov_local_2021}, including the shift of the second PDF peak towards lower $r$-values at lower temperatures. However, we present data that goes beyond previous reports by showing a) the appearance of peak splitting in $G(r)$ peaks, b) collecting data at temperatures down to 95 K, where the peak splitting becomes more distinct, and c) the persistence of monoclinic phase well up to the low temperature. The persistence of the monoclinic phase at all temperatures during both cooling and warming cycles without any major structural changes was concluded by analyzing results from the PDF experiment using small box modeling implemented in PDFGui routines. Ref. \cite{petkov_local_2021} observed the presence of a mixed stacking with at least 50\% monoclinic stacking persisting down to 150K, notably lower than 100\% at 300K. However, the quality of fits using single phases (1$T$$'$ or $T_d$) are not very different; this does not rule out the possibility of the persistence of the monoclinic phase down to lower temperatures, such as in this study. 

The structure's temperature profile might also depend on sample history, as suggested by several previous studies on MoTe$_2$. \cite{petkov_local_2021,heikes_mechanical_2018} The sharp first peak on the PDF, corresponding to the Mo-Te bond, changed insignificantly during thermal cycles and indicated the presence of strong atom-atom correlations. Strong atomic correlation, in addition to the static and dynamic disorders in MoTe$_2$, may be the cause of the suppression of the SPT, persistence of phases, and the local structure memory effects. Local strains that frustrate the long-range octahedral coordination environment have been shown to completely suppress the SPT in metal-halides perovskites \cite{van_de_goor_local_2022} while epitaxial strain and vacancies in oxides such as VO$_2$ \cite{kudo_suppression_2013,hoshi_structural_2020}. Complete suppression of SPT as an effect of correlated motion (dynamic disorder) has been observed in Ge$_2$Sb$_2$Te$_5$. \cite{qi_photoinduced_2022}

Small box modeling provided insights into the temperature evolution of atomic correlation in topological phases of MoTe$_2$. However, only limited information on the atomic correlations can be obtained by this technique since the deviation of the mean square relative displacement of atoms extracted from the PDF peak widths ($\sigma_{ij}$) is insensitive to the phonon density of states. \cite{jeong_lattice_2003} Large box modeling of the PDF data provides more information on the pair correlations and the effects of low-frequency phonon modes, comparable to techniques such as inelastic neutron scattering (INS). \cite{goodwin_model-independent_2005} A large box modeling refinement using RMCProfile indicated a preferential motion of Te atoms along the stacking direction. This was more distinct at higher temperatures. At elevated temperatures, a higher mean-squared relative displacement (MSRD) for atomic pairs corresponding to the interlayer distances has been observed in layered 2$H$-MoS$_2$. \cite{pudza_unraveling_2023} This implies that the chalcogen atoms have non-isotropic $U_{iso}$ with $U_{33}$ values that are larger than other $U_{iso}$ values. Such specific motions have also been observed for oxygen atoms in CuMn$_2$O$_4$ and A site disorders in Bi$_{0.5}$K$_{0.5}$TiO$_3$ (BKT), \cite{shoemaker_unraveling_2009,jiang_local_2017} and is likely due to the small distortion of the Mo-Te octahedra in MoTe$_2$. The atomic interactions across the vdW gap are weak, leading to a smaller distortion within the layer. A more rigorous study, including a detailed investigation of phonon spectrums of the temperature dependence of correlated motion in topological MoTe$_2$ by using advanced modeling \cite{dove_review_2022,goodwin_model-independent_2005}, can provide additional insights into the temperature-dependent dynamical processes of the local structure in MoTe$_2$. 

\section{CONCLUSION}
Average and short-range structural properties of MoTe$_2$ were studied using synchrotron XRD and PDF techniques. Synchrotron XRD showed clear signs of the appearance of the orthorhombic $T_d$ phase at lower temperatures; however, multiphase Rietveld refinement showed the phase fraction of the orthorhombic phase to be $\sim$0.25  even at room temperature. The monoclinic phase also persisted down to the lowest measured temperature (95 K), indicating suppression of structural phase transition and memory effects during thermal cycles for both average and local structure analysis. As discussed, this is most likely due to static and dynamic disorders. Small box modeling of the PDF data showed influences of strong atom-atom correlations, and large box modeling further revealed a preferential motion of Te atoms towards the $c$-axis at all higher temperatures. Our results provide a strong foundation to investigate the influences of disorders and atomic correlations on the global and local SPT using advanced scattering techniques in MoTe$_2$ and other TMDs. A more detailed study of correlated motion in topological MoTe$_2$ by considering phonon contributions to the PDF modeling may provide more insights regarding the dynamical processes involving the temperature-dependent local structure in MoTe$_2$. \cite{dove_review_2022,goodwin_model-independent_2005} Detailed information on the correlated motion can also be obtained via the Extended X-ray Absorption Fine Structure (EXAFS) technique. The specificity and sensitivity of EXAFS are, in contrast to the PDF technique, suitable for studying the correlated motion and its effects on the local structure.

\begin{acknowledgments}
The authors acknowledge support from the U.S. Air Force Oﬀice of Scientific Research and Clarkson Aerospace Corp. under Award FA9550-21-1-0460. B.F. acknowledges Welch Professorship support from the Welch Foundation (Grant No. L-E-001- 19921203), support from the Texas Center for Superconductivity at the University of Houston (TcSUH), and the Hewlett Packard Enterprise Data Science Institute for providing resources for high-performance scientific computations. B.F and S.K. thank M. Abeykoon for the preliminary PDF experiments at NSLS-II and Y.C. Shao for insights into the EXAFS experiment. B.F. and S.K. also thank A. Alfailakawi and B. Karki for their suggestions on the experimental planning and preliminary data analysis. This research used resources of the Advanced Photon Source, a U.S. Department of Energy (DOE) Oﬀice of Science user facility operated for the DOE Oﬀice of Science by Argonne National Laboratory under Contract No. DE-AC02-06CH11357. 
\end{acknowledgments}

\clearpage
\bibliographystyle{unsrt} 
\bibliography{ms}

\begin{thebibliography}{10}

\bibitem{jariwala_emerging_2014}
Deep Jariwala, Vinod~K. Sangwan, Lincoln~J. Lauhon, Tobin~J. Marks, and Mark~C. Hersam.
\newblock Emerging {device} {applications} for {semiconducting} {two}-{dimensional} {transition} {metal} {dichalcogenides}.
\newblock {\em ACS Nano}, 8(2):1102--1120, February 2014.

\bibitem{tamai_fermi_2016}
A.~Tamai, Q.~S. Wu, I.~Cucchi, F.~Y. Bruno, S.~Riccò, T.~K. Kim, M.~Hoesch, C.~Barreteau, E.~Giannini, C.~Besnard, A.~A. Soluyanov, and F.~Baumberger.
\newblock Fermi arcs and {their} {topological} {character} in the {candidate} {Type}-{II} {Weyl} {semimetal} $\mathrm{MoTe}_2$.
\newblock {\em Phys. Rev. X}, 6(3):031021, August 2016.

\bibitem{dahal_tunable_2020}
R.~Dahal, L.~Z. Deng, N.~Poudel, M.~Gooch, Z.~Wu, H.~C. Wu, H.~D. Yang, C.~K. Chang, and C.~W. Chu.
\newblock Tunable structural phase transition and superconductivity in the {Weyl} semimetal $\mathrm{Mo}_{1-x}\mathrm{W}_{x}\mathrm{Te}_{2}$.
\newblock {\em Phys. Rev. B}, 101(14):140505, April 2020.

\bibitem{heikes_mechanical_2018}
Colin Heikes, I-Lin Liu, Tristin Metz, Chris Eckberg, Paul Neves, Yan Wu, Linda Hung, Phil Piccoli, Huibo Cao, Juscelino Leao, Johnpierre Paglione, Taner Yildirim, Nicholas~P. Butch, and William Ratcliff.
\newblock Mechanical control of crystal symmetry and superconductivity in {Weyl} semimetal $\mathrm{{MoTe}_{2}}$.
\newblock {\em Phys. Rev. Mater.}, 2(7):074202, July 2018.

\bibitem{zhang_light-induced_2019}
M.Y. Zhang, Z.X. Wang, Y.N. Li, L.Y. Shi, D.~Wu, T.~Lin, S.J. Zhang, Y.Q. Liu, Q.M. Liu, J.~Wang, T.~Dong, and N.L. Wang.
\newblock Light-{induced} {subpicosecond} {lattice} {symmetry} {switch} in $\mathrm{MoTe}_2$.
\newblock {\em Phys. Rev. X}, 9(2):021036, May 2019.

\bibitem{fathipour_exfoliated_2014}
S.~Fathipour, N.~Ma, W.~S. Hwang, V.~Protasenko, S.~Vishwanath, H.~G. Xing, H.~Xu, D.~Jena, J.~Appenzeller, and A.~Seabaugh.
\newblock Exfoliated multilayer $\mathrm{MoTe}_2$ field-effect transistors.
\newblock {\em Appl. Phys. Lett.}, 105(19):192101, November 2014.

\bibitem{iqbal_tailoring_2019}
M.~W. Iqbal, Aliya Amin, M.~A. Kamran, Hira Ateeq, Ehsan Elahi, G.~Hussain, Sikander Azam, Sikandar Aftab, Thamer Alharbi, and Abdul Majid.
\newblock Tailoring the electrical properties of $\mathrm{MoTe}_2$ field effect transistor via chemical doping.
\newblock {\em Superlattice Microst.}, 135:106247, November 2019.

\bibitem{liu_atomic_2021}
Xia Liu, Arnob Islam, Ning Yang, Bradley Odhner, Mary~Anne Tupta, Jing Guo, and Philip X.-L. Feng.
\newblock Atomic {layer} $\mathrm{MoTe_{2}}$ {Field}-{Effect} {transistors} and {monolithic} {logic} {circuits} {configured} by {scanning} {laser} {annealing}.
\newblock {\em ACS Nano}, 15(12):19733--19742, December 2021.

\bibitem{qi_superconductivity_2016}
Yanpeng Qi, Pavel~G. Naumov, Mazhar~N. Ali, Catherine~R. Rajamathi, Walter Schnelle, Oleg Barkalov, Michael Hanfland, Shu-Chun Wu, Chandra Shekhar, Yan Sun, Vicky Süß, Marcus Schmidt, Ulrich Schwarz, Eckhard Pippel, Peter Werner, Reinald Hillebrand, Tobias Förster, Erik Kampert, Stuart Parkin, R.~J. Cava, Claudia Felser, Binghai Yan, and Sergey~A. Medvedev.
\newblock Superconductivity in {Weyl} semimetal candidate $\mathrm{MoTe}_2$.
\newblock {\em Nat Commun}, 7(1):11038, March 2016.

\bibitem{kim_origins_2017}
Hyun-Jung Kim, Seoung-Hun Kang, Ikutaro Hamada, and Young-Woo Son.
\newblock Origins of the structural phase transitions in $\mathrm{MoTe}_{2}$ and $\mathrm{WTe}_{2}$.
\newblock {\em Phys. Rev. B}, 95(18):180101, May 2017.

\bibitem{yang_structural_2017}
Heejun Yang, Sung~Wng Kim, Manish Chhowalla, and Young~Hee Lee.
\newblock Structural and quantum-state phase transitions in van der {Waals} layered materials.
\newblock {\em Nature Phys}, 13(10):931--937, October 2017.

\bibitem{sun_prediction_2015}
Yan Sun, Shu-Chun Wu, Mazhar~N. Ali, Claudia Felser, and Binghai Yan.
\newblock Prediction of {Weyl} semimetal in orthorhombic $\mathrm{{MoTe}_{2}}$.
\newblock {\em Phys. Rev. B}, 92(16):161107, October 2015.

\bibitem{hughes_electrical_1978}
H.~P. Hughes and R.~H. Friend.
\newblock Electrical resistivity anomaly in $\mathrm{{\beta}-{MoTe}_{2}}$ (metallic behaviour).
\newblock {\em J. Phys. C: Solid State Phys.}, 11(3):L103, February 1978.

\bibitem{yang_elastic_2017}
Junjie Yang, Jonathan Colen, Jun Liu, Manh~Cuong Nguyen, Gia-wei Chern, and Despina Louca.
\newblock Elastic and electronic tuning of magnetoresistance in $\mathrm{{MoTe}_{2}}$.
\newblock {\em Sci. Adv.}, 3(12):eaao4949, December 2017.

\bibitem{clarke_low-temperature_1978}
R.~Clarke, E.~Marseglia, and H.~P. Hughes.
\newblock A low-temperature structural phase transition in $\mathrm{{\beta}-{MoTe}_{2}}$.
\newblock {\em Philos. Mag. B}, 38(2):121--126, August 1978.

\bibitem{karki_strain-induced_2020}
Bhupendra Karki, Byron Freelon, Manthila Rajapakse, Rajib Musa, S.~M.~Shah Riyadh, Blake Morris, Usman Abu, Ming Yu, Gamini Sumanasekera, and Jacek~B. Jasinski.
\newblock Strain-induced vibrational properties of few layer black phosphorus and $\mathrm{{MoTe}_{2}}$ {via} {Raman} spectroscopy.
\newblock {\em Nanotechnology}, 31(42):425707, July 2020.

\bibitem{sie_ultrafast_2019}
Edbert~J. Sie, Clara~M. Nyby, C.~D. Pemmaraju, Su~Ji Park, Xiaozhe Shen, Jie Yang, Matthias~C. Hoffmann, B.~K. Ofori-Okai, Renkai Li, Alexander~H. Reid, Stephen Weathersby, Ehren Mannebach, Nathan Finney, Daniel Rhodes, Daniel Chenet, Abhinandan Antony, Luis Balicas, James Hone, Thomas~P. Devereaux, Tony~F. Heinz, Xijie Wang, and Aaron~M. Lindenberg.
\newblock An ultrafast symmetry switch in a {Weyl} semimetal.
\newblock {\em Nature}, 565(7737):61--66, January 2019.

\bibitem{kuiri_thickness_2020}
Manabendra Kuiri, Subhadip Das, D.~V.~S. Muthu, Anindya Das, and A.~K. Sood.
\newblock Thickness dependent transition from the $\mathrm{{1T'}}$ to {Weyl} semimetal phase in ultrathin $\mathrm{{MoTe}_{2}}$: electrical transport, noise and {Raman} studies.
\newblock {\em Nanoscale}, 12(15):8371--8378, April 2020.

\bibitem{schneeloch_emergence_2019}
John~A. Schneeloch, Chunruo Duan, Junjie Yang, Jun Liu, Xiaoping Wang, and Despina Louca.
\newblock Emergence of topologically protected states in the $\mathrm{MoTe}_2$ {Weyl} semimetal with layer-stacking order.
\newblock {\em Phys. Rev. B}, 99(16):161105, April 2019.

\bibitem{qi_traversing_2022}
Yingpeng Qi, Mengxue Guan, Daniela Zahn, Thomas Vasileiadis, Hélène Seiler, Yoav~William Windsor, Hui Zhao, Sheng Meng, and Ralph Ernstorfer.
\newblock Traversing {double}-{well} {potential} {energy} {surfaces}: {photoinduced} {concurrent} {intralayer} and {interlayer} {structural} {transitions} in $\mathrm{{XTe}_{2}}$ ({X} = {Mo}, {W}).
\newblock {\em ACS Nano}, 16(7):11124--11135, July 2022.

\bibitem{singh_engineering_2020}
Sobhit Singh, Jinwoong Kim, Karin~M. Rabe, and David Vanderbilt.
\newblock Engineering {weyl} {phases} and {nonlinear} {Hall} {effects} in $\mathrm{T}_d$ $\mathrm{MoTe}_2$.
\newblock {\em Phys. Rev. Lett.}, 125(4):046402, July 2020.

\bibitem{caramazza_temperature_2016}
S.~Caramazza, C.~Marini, L.~Simonelli, P.~Dore, and P.~Postorino.
\newblock Temperature dependent {EXAFS} study on transition metal dichalcogenides $\mathrm{{MoX}_{2}}$ ({X} = {S}, {Se}, {Te}).
\newblock {\em J. Phys.: Condens. Matter}, 28(32):325401, June 2016.

\bibitem{petkov_local_2021}
Valeri Petkov and Yang Ren.
\newblock Local structure memory effects in the polar and nonpolar phases of $\mathrm{{MoTe}_{2}}$.
\newblock {\em Phys. Rev. B}, 103(9):094101, March 2021.

\bibitem{philip_out--plane_2021}
Sharon~S. Philip, Anushika Athauda, Yosuke Goto, Yoshikazu Mizuguchi, and Despina Louca.
\newblock Out-of-{plane} {sulfur} {distortions} in the $\mathrm{{Bi}_4\mathrm{O}_4\mathrm{S}3}$ {superconductor}.
\newblock {\em Condens. Matter}, 6(4):48, November 2021.

\bibitem{smith_crystal_2008}
Millicent~B. Smith, Katharine Page, Theo Siegrist, Peter~L. Redmond, Erich~C. Walter, Ram Seshadri, Louis~E. Brus, and Michael~L. Steigerwald.
\newblock Crystal {structure} and the {paraelectric}-to-{ferroelectric} {phase} {transition} of {nanoscale} $\mathrm{BaTiO}_3$.
\newblock {\em J. Am. Chem. Soc.}, 130(22):6955--6963, June 2008.

\bibitem{subires_order-disorder_2023}
D.~Subires, A.~Korshunov, A.~H. Said, L.~Sánchez, Brenden~R. Ortiz, Stephen~D. Wilson, A.~Bosak, and S.~Blanco-Canosa.
\newblock Order-disorder charge density wave instability in the kagome metal ($\mathrm{Cs},\mathrm{Rb})\mathrm{V}_3\mathrm{Sb}_5$.
\newblock {\em Nat Commun}, 14(1):1015, February 2023.

\bibitem{malliakas_square_2005}
Christos Malliakas, Simon J.~L. Billinge, Hyun~Jeong Kim, and Mercouri~G. Kanatzidis.
\newblock Square {nets} of {tellurium}: {rare}-{earth} {dependent} {variation} in the {charge density} {wave} of $\mathrm{RETe}_3$ ({RE} = {rare}-{earth} {element}).
\newblock {\em J. Am. Chem. Soc.}, 127(18):6510--6511, May 2005.

\bibitem{kim_local_2006}
H.~J. Kim, C.~D. Malliakas, A.~Tomic, S.~H. Tessmer, M.~G. Kanatzidis, and S.~J.~L. Billinge.
\newblock Local atomic structure and discommensurations in the charge density wave of $\mathrm{CeTe}_3$.
\newblock {\em Phys. Rev. Lett.}, 96(22):226401, June 2006.

\bibitem{noauthor_mote2_nodate}
{MoTe$\mathrm{_{2}}$} {1$\mathrm{T'}$} - {Molybdenum} {Ditelluride}, {See} https://www.hqgraphene.com/{1T-MoTe2}.php.

\bibitem{toby_gsas-ii_2013}
Brian~H. Toby and Robert~B. Von~Dreele.
\newblock \textit{{GSAS}}-{II} : the genesis of a modern open-source all purpose crystallography software package.
\newblock {\em J Appl Crystallogr}, 46(2):544--549, April 2013.

\bibitem{farrow_pdffit2_2007}
C.~L. Farrow, P.~Juhas, J.~W. Liu, D.~Bryndin, E.~S. Božin, J.~Bloch, Th~Proffen, and S.~J.~L. Billinge.
\newblock {PDFfit2} and {PDFgui}: computer programs for studying nanostructure in crystals.
\newblock {\em J. Phys.: Condens. Matter}, 19(33):335219, July 2007.

\bibitem{juhas_pdfgetx3_2013}
P.~Juhás, T.~Davis, C. L. Farrow, and S. J. L. Billinge.
\newblock \textit{{PDFgetX3}} : a rapid and highly automatable program for processing powder diffraction data into total scattering pair distribution functions.
\newblock {\em J Appl Crystallogr}, 46(2):560--566, April 2013.

\bibitem{pourpoint_new_2012}
Frédérique Pourpoint, Xiao Hua, Derek~S. Middlemiss, Paul Adamson, Da~Wang, Peter~G. Bruce, and Clare~P. Grey.
\newblock New {insights} into the {crystal} and {electronic} {structures} of $\mathrm{Li}_{1+x}\mathrm{V}_{1–x}\mathrm{O}_2$ from {solid} {state} {NMR}, {pair} {distribution} {function} {analyses}, and {first} {principles} {calculations}.
\newblock {\em Chem. Mater.}, 24(15):2880--2893, August 2012.

\bibitem{rietveld_rietveld_2014}
Hugo~M Rietveld.
\newblock The {Rietveld} method.
\newblock {\em Phys. Scr.}, 89(9):098002, September 2014.

\bibitem{runcevski_rietveld_2021}
Tomče Runčevski and Craig~M. Brown.
\newblock The {Rietveld} {Refinement} {method}: {half} of a {century} {anniversary}.
\newblock {\em Cryst. Growth Des.}, 21(9):4821--4822, September 2021.

\bibitem{tao_appearance_2019}
Yu~Tao, John~A. Schneeloch, Chunruo Duan, Masaaki Matsuda, Sachith~E. Dissanayake, Adam~A. Aczel, Jaime~A. Fernandez-Baca, Feng Ye, and Despina Louca.
\newblock Appearance of a $\mathrm{{T}_{d}^{*}}$ phase across the $\mathrm{{T}_{d} \rightarrow 1{T'}}$ phase boundary in the {Weyl} semimetal $\mathrm{{MoTe}_{2}}$.
\newblock {\em Phys. Rev. B}, 100(10):100101, September 2019.

\bibitem{hovden_atomic_2016}
Robert Hovden, Adam~W. Tsen, Pengzi Liu, Benjamin~H. Savitzky, Ismail El~Baggari, Yu~Liu, Wenjian Lu, Yuping Sun, Philip Kim, Abhay~N. Pasupathy, and Lena~F. Kourkoutis.
\newblock Atomic lattice disorder in charge-density-wave phases of exfoliated dichalcogenides ({1T}-$\mathrm{TaS_{2}}$).
\newblock {\em Proc Natl Acad Sci USA}, 113(41):11420--11424, October 2016.

\bibitem{yan_composition_2017}
Xue-Jun Yan, Yang-Yang Lv, Lei Li, Xiao Li, Shu-Hua Yao, Yan-Bin Chen, Xiao-Ping Liu, Hong Lu, Ming-Hui Lu, and Yan-Feng Chen.
\newblock Composition dependent phase transition and its induced hysteretic effect in the thermal conductivity of $\mathrm{W_{x}Mo_{1-x}Te_{2}}$.
\newblock {\em Appl. Phys. Lett.}, 110(21):211904, May 2017.

\bibitem{lee_origin_2019}
Sung-Hoon Lee, Jung~Suk Goh, and Doohee Cho.
\newblock Origin of the {insulating} {phase} and {first}-{order} {metal}-{insulator} {transition} in {1T}-$\mathrm{{TaS}_{2}}$.
\newblock {\em Phys. Rev. Lett.}, 122(10):106404, March 2019.

\bibitem{hart_emergence_2022}
James~L. Hart, Lopa Bhatt, Yanbing Zhu, Myung-Geun Han, Elisabeth Bianco, Shunran Li, David~J. Hynek, John~A. Schneeloch, Yu~Tao, Despina Louca, Peijun Guo, Yimei Zhu, Felipe Jornada, Evan~J. Reed, Lena~F. Kourkoutis, and Judy~J. Cha.
\newblock Emergence of {layer} {stacking} {disorder} in c-axis {confined} {MoTe}$_2$, October 2022.

\bibitem{cheon_structural_2021}
Yeryun Cheon, Soo~Yeon Lim, Kangwon Kim, and Hyeonsik Cheong.
\newblock Structural phase transition and interlayer coupling in few-layer $\mathrm{{1T'}}$ and $\mathrm{T_{d}}$ $\mathrm{MoTe}_2$.
\newblock {\em ACS Nano}, 15(2):2962--2970, February 2021.

\bibitem{schneeloch_antiferromagnetic-ferromagnetic_2023}
John~A. Schneeloch, Luke Daemen, and Despina Louca.
\newblock Antiferromagnetic-ferromagnetic homostructures with {Dirac} magnons in van der {Waals} magnet {CrI}$_3$, May 2023.

\bibitem{egami_local_2004}
Takeshi Egami.
\newblock Local crystallography of crystals with disorder.
\newblock {\em Z. Krist. - Cryst. Mater.}, 219(3):122--129, March 2004.

\bibitem{billinge_rise_2019}
Simon J.~L. Billinge.
\newblock The rise of the {X}-ray atomic pair distribution function method: a series of fortunate events.
\newblock {\em Philos. T. R. Soc. A}, 377(2147):20180413, April 2019.

\bibitem{dove_review_2022}
Martin~T. Dove and Gong Li.
\newblock Review: {Pair} distribution functions from neutron total scattering for the study of local structure in disordered materials.
\newblock {\em Nuclear Analysis}, 1(4):100037, December 2022.

\bibitem{khadka_assessment_nodate}
Sumit Khadka, Gallington Leighhanne, and Byron Freelon.
\newblock An {Assessment} of {thermally} {driven} {local} {structural} {phase} {changes} in $\mathrm{{1T'}-{MoTe_{2}}}$ {Supplemental} {Material} ({S}.{M}.).

\bibitem{jeong_lattice_2003}
I.-K. Jeong, R.~H. Heffner, M.~J. Graf, and S.~J.~L. Billinge.
\newblock Lattice dynamics and correlated atomic motion from the atomic pair distribution function.
\newblock {\em Phys. Rev. B}, 67(10):104301, March 2003.

\bibitem{kim_thermomechanical_2021}
Dohyun Kim, Jun-Ho Lee, Kyungrok Kang, Dongyeun Won, Min Kwon, Suyeon Cho, Young-Woo Son, and Heejun Yang.
\newblock Thermomechanical {manipulation} of {electric} {transport} in $\mathrm{{MoTe}_{2}}$.
\newblock {\em Adv. Electron Mater.}, 7(4):2000823, 2021.

\bibitem{tucker_rmcprofile_2007}
Matthew~G Tucker, David~A Keen, Martin~T Dove, Andrew~L Goodwin, and Qun Hui.
\newblock {RMCProfile}: reverse {Monte} {Carlo} for polycrystalline materials.
\newblock {\em J. Phys.: Condens. Matter}, 19(33):335218, August 2007.

\bibitem{tan_rigid_2023}
Lei Tan, Volker Heine, Gong Li, and Martin~T Dove.
\newblock The {Rigid} {Unit} {Mode} model: review of ideas and applications.
\newblock {\em Rep. Prog. Phys.}, 2023.

\bibitem{tao_role_2003}
J.~Z. Tao and A.~W. Sleight.
\newblock The role of rigid unit modes in negative thermal expansion.
\newblock {\em J. Solid State Chem}, 173(2):442--448, 2003.

\bibitem{dove_which_2023}
Martin~T. Dove, Zhongsheng Wei, Anthony~E. Phillips, David~A. Keen, and Keith Refson.
\newblock Which phonons contribute most to negative thermal expansion in $\mathrm{ScF_{3}}$?
\newblock {\em APL Materials}, 11(4):041130, April 2023.

\bibitem{paul_tailoring_2020}
Suvodeep Paul, Saheb Karak, Manasi Mandal, Ankita Ram, Sourav Marik, R.~P. Singh, and Surajit Saha.
\newblock Tailoring the phase transition and electron-phonon coupling in $1\mathrm{T'}$-$\mathrm{MoTe}_2$ by charge doping: {A} {Raman} study.
\newblock {\em Phys. Rev. B}, 102(5):054103, August 2020.

\bibitem{he_dimensionality-driven_2018}
Rui He, Shazhou Zhong, Hyun~Ho Kim, Gaihua Ye, Zhipeng Ye, Logan Winford, Daniel McHaffie, Ivana Rilak, Fangchu Chen, Xuan Luo, Yuping Sun, and Adam~W. Tsen.
\newblock Dimensionality-driven orthorhombic $\mathrm{{MoTe}_{2}}$ at room temperature.
\newblock {\em Phys. Rev. B}, 97(4):041410, January 2018.

\bibitem{mandal_enhancement_2018}
Manasi Mandal, Sourav Marik, K.~P. Sajilesh, {Arushi}, Deepak Singh, Jayita Chakraborty, Nirmal Ganguli, and R.~P. Singh.
\newblock Enhancement of the superconducting transition temperature by {Re} doping in {Weyl} semimetal $\mathrm{{MoTe}_{2}}$.
\newblock {\em Phys. Rev. Mater.}, 2(9):094201, September 2018.

\bibitem{sakai_critical_2016}
Hideaki Sakai, Koji Ikeura, Mohammad~Saeed Bahramy, Naoki Ogawa, Daisuke Hashizume, Jun Fujioka, Yoshinori Tokura, and Shintaro Ishiwata.
\newblock Critical enhancement of thermopower in a chemically tuned polar semimetal $\mathrm{{MoTe}_{2}}$.
\newblock {\em Sci. Adv.}, 2(11):e1601378, November 2016.

\bibitem{van_de_goor_local_2022}
Tim Van De~Goor.
\newblock Local structure of hybrid metal-halide perovskites.
\newblock {\em Apollo - University of Cambridge Repository}, July 2022.

\bibitem{kudo_suppression_2013}
Kazutaka Kudo, Masakazu Kobayashi, Sunseng Pyon, and Minoru Nohara.
\newblock Suppression of structural {phase} {transition} in $\mathrm{{IrTe}_{2}}$ by {isovalent} {Rh} {doping}.
\newblock {\em J. Phys. Soc. Jpn.}, 82(8):085001, August 2013.

\bibitem{hoshi_structural_2020}
Kazuhisa Hoshi, Shunsuke Sakuragi, Takeshi Yajima, Yosuke Goto, Akira Miura, Chikako Moriyoshi, Yoshihiro Kuroiwa, and Yoshikazu Mizuguchi.
\newblock Structural {phase} {diagram} of $\mathrm{LaO}_{1-x}\mathrm{F}_x\mathrm{BiSSe}$: {suppression} of the {structural} {phase} {transition} by {partial} {F} {substitutions}.
\newblock {\em Condens. Matter}, 5(4):81, December 2020.

\bibitem{qi_photoinduced_2022}
Yingpeng Qi, Nianke Chen, Thomas Vasileiadis, Daniela Zahn, Hélène Seiler, Xianbin Li, and Ralph Ernstorfer.
\newblock Photoinduced ultrafast transition of the local correlated structure in chalcogenide phase-change materials.
\newblock {\em Phys. Rev. Lett.}, 129(13):135701, September 2022.

\bibitem{goodwin_model-independent_2005}
Andrew~L. Goodwin, Matthew~G. Tucker, Elizabeth~R. Cope, Martin~T. Dove, and David~A. Keen.
\newblock Model-independent extraction of dynamical information from powder diffraction data.
\newblock {\em Phys. Rev. B}, 72(21):214304, December 2005.

\bibitem{pudza_unraveling_2023}
Inga Pudza, Dmitry Bocharov, Andris Anspoks, Matthias Krack, Aleksandr Kalinko, Edmund Welter, and Alexei Kuzmin.
\newblock Unraveling the interlayer and intralayer coupling in two-dimensional layered {MoS2} by {X}-ray absorption spectroscopy and ab initio molecular dynamics simulations.
\newblock {\em Materials Today Communications}, 35:106359, June 2023.

\bibitem{shoemaker_unraveling_2009}
Daniel~P. Shoemaker, Jun Li, and Ram Seshadri.
\newblock Unraveling {atomic} {positions} in an {oxide} {spinel} with {two} {Jahn-}{Teller} {ions}: {local} {structure} {investigation} of $\mathrm{CuMn}_2\mathrm{O}_4$.
\newblock {\em J. Am. Chem. Soc.}, 131(32):11450--11457, August 2009.

\bibitem{jiang_local_2017}
Bo~Jiang, Tor Grande, and Sverre~M. Selbach.
\newblock Local {structure} of {disordered} $\mathrm{Bi}_{0.5}\mathrm{K}_{0.5}\mathrm{TiO}_3$ {investigated} by {pair} {distribution} {function} {analysis} and {first}-{principles} {calculations}.
\newblock {\em Chem. Mater.}, 29(10):4244--4252, May 2017.

\end{thebibliography}


\begin{thebibliography}{10}

\bibitem{egami_local_2004}
Takeshi Egami.
\newblock Local crystallography of crystals with disorder.
\newblock {\em Z. Krist. - Cryst. Mater.}, 219(3):122--129, March 2004.

\bibitem{billinge_rise_2019}
Simon J.~L. Billinge.
\newblock The rise of the {X}-ray atomic pair distribution function method: a series of fortunate events.
\newblock {\em Philos. T. R. Soc. A}, 377(2147):20180413, April 2019.

\bibitem{gebretsadik_study_2023}
Adane Gebretsadik, Ruizhe Wang, Arwa Alyami, Hind Adawi, Jean-Guy Lussier, Katharine~L. Page, and Almut Schroeder.
\newblock Study of atomic disorder in {Ni}-{V} alloys, February 2023.

\bibitem{farrow_pdffit2_2007}
C.~L. Farrow, P.~Juhas, J.~W. Liu, D.~Bryndin, E.~S. Božin, J.~Bloch, Th~Proffen, and S.~J.~L. Billinge.
\newblock {PDFfit2} and {PDFgui}: computer programs for studying nanostructure in crystals.
\newblock {\em J. Phys.: Condens. Matter}, 19(33):335219, July 2007.

\bibitem{juhas_complex_2015}
P.~Juhás, C.~Farrow, X.~Yang, K.~Knox, and S.~Billinge.
\newblock Complex modeling: a strategy and software program for combining multiple information sources to solve ill posed structure and nanostructure inverse problems.
\newblock {\em Acta Cryst A}, 71(6):562--568, November 2015.

\bibitem{coelho_topas_2018}
Alan~A. Coelho.
\newblock \textit{{TOPAS}} and \textit{{TOPAS}-{Academic}} : an optimization program integrating computer algebra and crystallographic objects written in {C}++.
\newblock {\em J Appl Crystallogr}, 51(1):210--218, February 2018.

\bibitem{tucker_rmcprofile_2007}
Matthew~G Tucker, David~A Keen, Martin~T Dove, Andrew~L Goodwin, and Qun Hui.
\newblock {RMCProfile}: reverse {Monte} {Carlo} for polycrystalline materials.
\newblock {\em J. Phys.: Condens. Matter}, 19(33):335218, August 2007.

\bibitem{toby_gsas-ii_2013}
Brian~H. Toby and Robert~B. Von~Dreele.
\newblock \textit{{GSAS}}-{II} : the genesis of a modern open-source all purpose crystallography software package.
\newblock {\em J Appl Crystallogr}, 46(2):544--549, April 2013.

\bibitem{rietveld_rietveld_2014}
Hugo~M Rietveld.
\newblock The {Rietveld} method.
\newblock {\em Phys. Scr.}, 89(9):098002, September 2014.

\bibitem{runcevski_rietveld_2021}
Tomče Runčevski and Craig~M. Brown.
\newblock The {Rietveld} {Refinement} {Method}: {half} of a {century} {anniversary}.
\newblock {\em Cryst. Growth Des.}, 21(9):4821--4822, September 2021.

\bibitem{tao_appearance_2019}
Yu~Tao, John~A. Schneeloch, Chunruo Duan, Masaaki Matsuda, Sachith~E. Dissanayake, Adam~A. Aczel, Jaime~A. Fernandez-Baca, Feng Ye, and Despina Louca.
\newblock Appearance of a $\mathrm{{T}_{d}^{*}}$ phase across the $\mathrm{{T}_{d} \rightarrow 1{T'}}$ phase boundary in the {Weyl} semimetal $\mathrm{{MoTe}_{2}}$.
\newblock {\em Phys. Rev. B}, 100(10):100101, September 2019.

\bibitem{schneeloch_emergence_2019}
John~A. Schneeloch, Chunruo Duan, Junjie Yang, Jun Liu, Xiaoping Wang, and Despina Louca.
\newblock Emergence of topologically protected states in the $\mathrm{MoTe}_2$ {Weyl} semimetal with layer-stacking order.
\newblock {\em Phys. Rev. B}, 99(16):161105, April 2019.

\end{thebibliography}

\end{document}


\title{An Assessment of Thermally Driven Local Structural Phase Changes in 1$T$$'$- MoTe$_2$ - Supplemental Material (S.M.)}

\author{S. Khadka}
\affiliation{Department of Physics and Texas Center for Superconductivity, University of Houston, Houston, Texas 77204, United States}
\author{L.C. Gallington}
\affiliation{X-ray Science Division, Advanced Photon Source, Argonne National Laboratory, Argonne, Illinois 60439-4858, United States}
\author{B. Freelon}
\affiliation{Department of Physics and Texas Center for Superconductivity, University of Houston, Houston, Texas 77204, United States}

\maketitle

\section{LOCAL STRUCTURE MODELING}
Diffraction experiments are commonly used to obtain the periodic structural arrangements in crystals. The method is entirely explained by Bragg's law $2dsin\theta = n\lambda$, where $d$ is the interatomic spacing, $\theta$ is the diffraction angle, $n$ is the order of diffraction, and $\lambda$ is the wavelength of the incident radiation. While highly successful in explaining the structural properties of crystalline materials, there are limitations when the Bragg expression is applied to crystals with a significant amount of disorders. \cite{egami_local_2004} For disordered systems, an approach that does not restrict periodicity and adopts non-crystallographic concepts for studying these disorders over short-distance scales is more effective. Using a powder diffraction-like experimental setup, the pair distribution function (PDF) technique is used to study the ordering that develops over these short distances. In this sense, PDF is a bulk probe sensitive to local structures. The technique has been proven effective in studying material properties that would otherwise be inaccessible to average structure (Bragg diffraction) experiments. \cite{billinge_rise_2019}

\begin{figure}[!hb]
    \centering
    \includegraphics[width=0.75\textwidth]{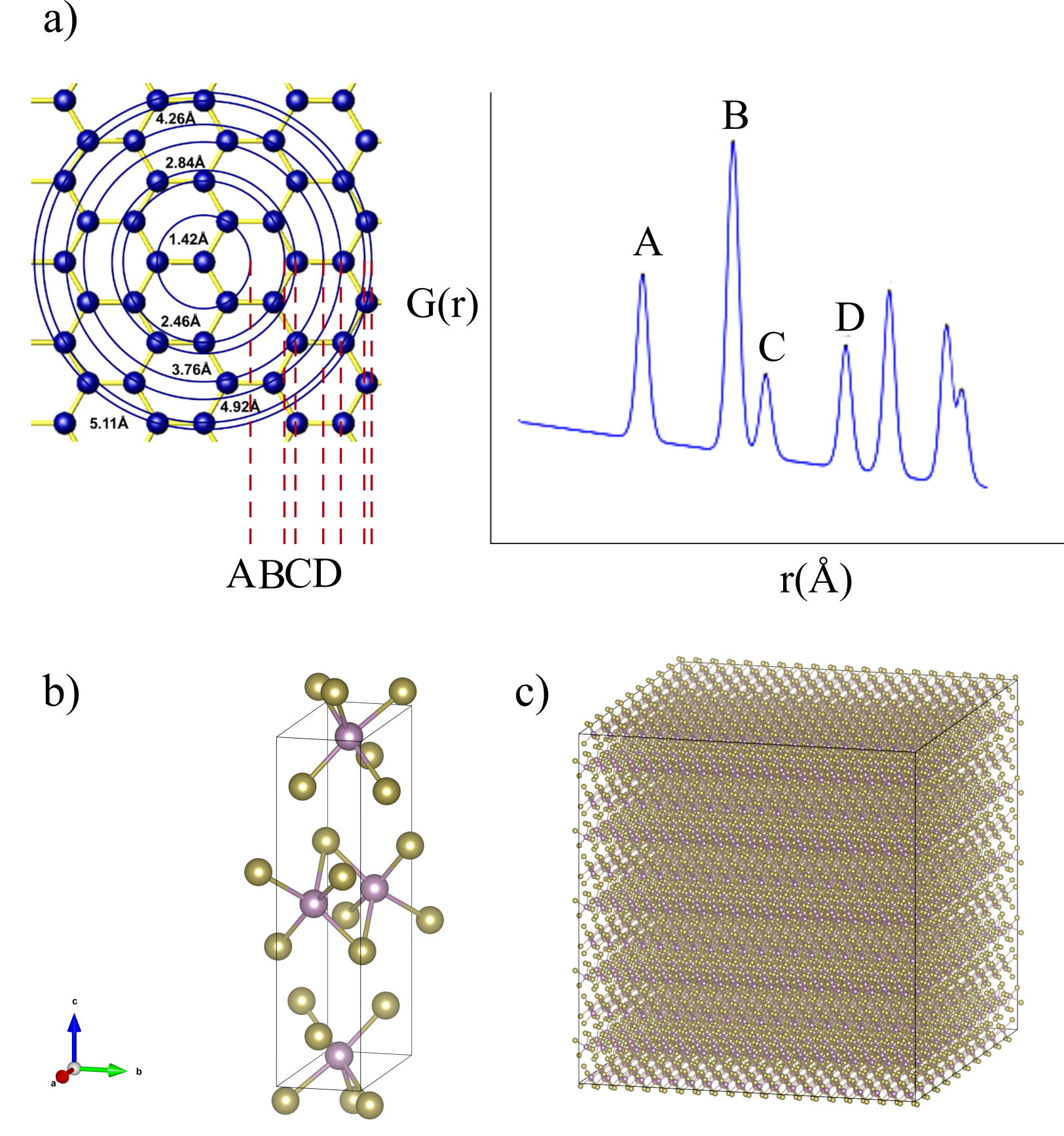}
    \caption{(a) $G(r)$ represents the real space interatomic distances and local coordination geometry. A, B, C, and D represent the interatomic distances corresponding to the first four peaks on the PDF curve. (b) An example of a unit cell used for small box modeling, which employs single or few unit cells with few tens of atoms for refinement, and (c) Example of a supercell used for large box modeling. The supercell usually contains thousands of atoms for refinement. A $10\times 20\times 5$ 1$T$' MoTe$_2$ supercell is shown here.}
    \label{smallandlargebox}
\end{figure}

The PDF is essentially the Fourier transform of the total scattering function, $S(Q)$, and can be expressed as
\begin{equation}
    G(r)=2/\pi\int_{Q_{min}}^{Q_{max}}[S(Q)-1]Qsin(Qr)dQ
\end{equation}
$Q_{min}$ and $Q_{max}$ are the instrument's minimum and maximum momentum transfer, respectively.
$S(Q)$ includes the normalized scattered intensity due to both Bragg scattering, non-Bragg diffuse scattering, and atomic correlations collected at the detector. $G(r)$ gives the probability of finding a pair of atoms separated by a distance $r$ in the material. \cite{gebretsadik_study_2023} As seen in FIG. \ref{smallandlargebox}(a) the atomic pair correlation function $G(r)$ directly represents the real space atomic pair distribution and gives information on the local coordination geometry of the system.

The analysis of PDF data is done by a refinement process, typically using two approaches: a) small box modeling and b) large box modeling. As shown in FIG. \ref{smallandlargebox}(b), a Rietveld-like refinement is performed in small box modeling using a single or a few unit cells with less than a hundred atoms. Lattice parameters, atomic positions, and ADPs are iteratively adjusted until a reasonable agreement is obtained between the calculated and experimental PDF data. A commonly used parameter to quantify the agreement is the weighted $R$-value (R$_w$) given by 
\begin{equation}
    R_w=\sqrt{\frac{\sum_{i=1}^N w(r_i)[G_{obs}(r_i)-G_{calc}(r_i)]^2}{\sum_{i=1}^N w(r_i)G_{obs}(r_i)^2}}
\end{equation}
where $G_{obs}$ is the experimental PDF, $G_{calc}$ is the calculated (simulated) PDF, and $w$ is the weight of each data point.\cite{farrow_pdffit2_2007} Programs like PDFGui, DiffpyCMI, and TOPAS are available for small-box modeling. \cite{farrow_pdffit2_2007,juhas_complex_2015,coelho_topas_2018}

In large box modeling, Reverse Monte Carlo (RMC) refinement is performed with a large supercell with atoms on the order of ten thousand, as shown in FIG. \ref{smallandlargebox}(c), to refine atomic structure based on total scattering data. \cite{tucker_rmcprofile_2007} An initial random arrangement of atoms is iteratively adjusted by changing their positions to match the experimentally obtained scattering profile. The simulation calculates the best-fit atomic configuration by minimizing the difference between the experimental data and the simulated patterns, resulting in an improved atomic model that can be used to extract the structural properties of the material under study. RMCProfile is a widely used software for large-box modeling.

\clearpage
\section{X-Ray diffraction data}
Merged peaks are observed at $2\theta$=2.954$^{\circ}$ and $2\theta$=3.236$^{\circ}$ as the temperature is lowered down to 95 K indexed as (112)$_\textrm{O}$ and (113)$_\textrm{O}$ for the orthorhombic, $T_d$ unit cell.

\begin{figure}[!ht]
    \centering
    \includegraphics[width=0.65\textwidth]{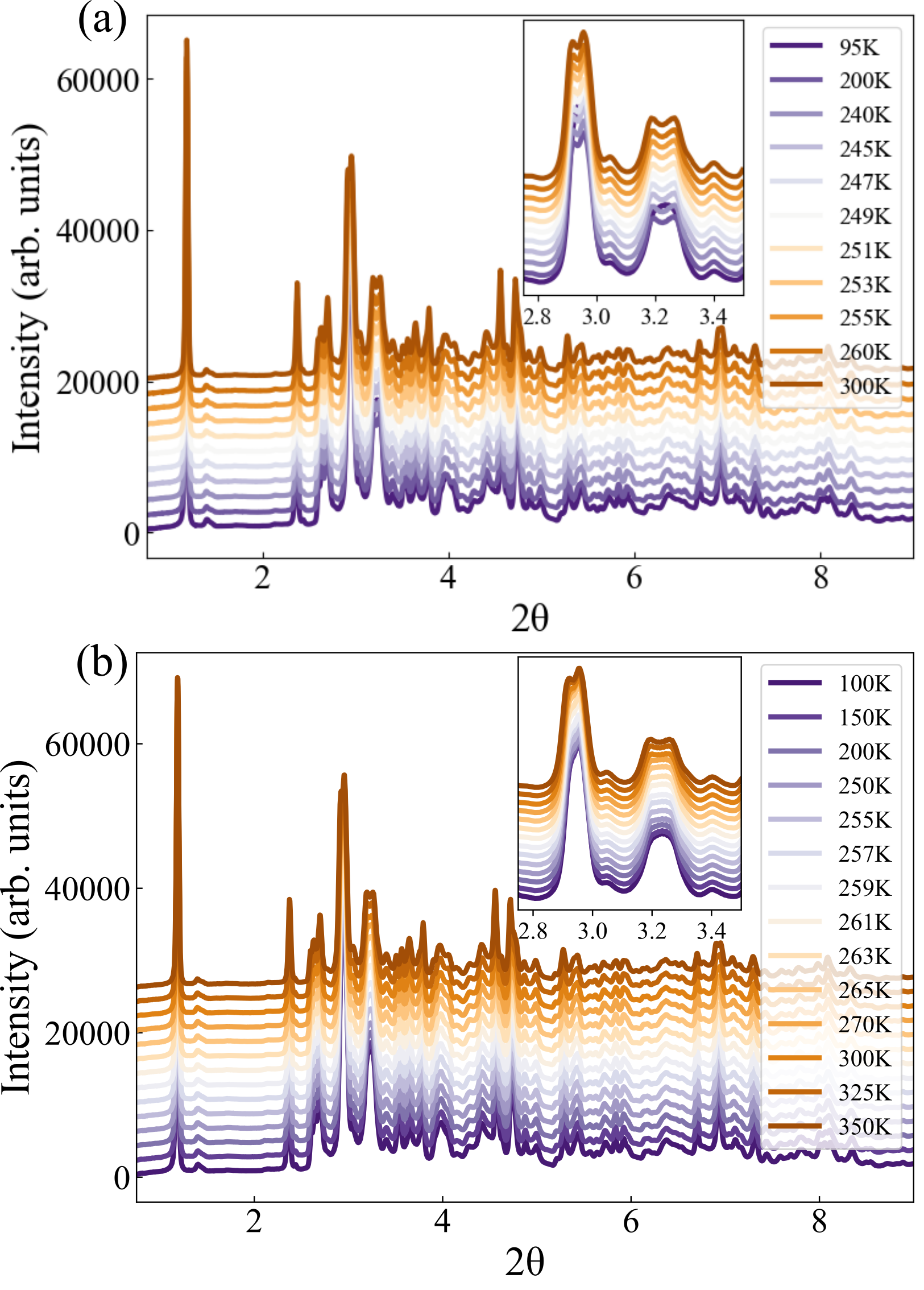}
    \caption{Powder X-ray diffraction patterns obtained during (a) cooling and (b) warming cycles during the experiment. The inset of (a) shows the splitting of peaks at higher temperatures, indicating the monoclinic phase. The splitting is not observed until the sample is warmed to room temperature during the warming cycle.}
    \label{fig:coolingXRD}    
\end{figure}

\clearpage
\section{Quality of fit for Rietveld refinements}
The quality of fit (figure of merit) was determined by comparing the Rietveld weighted profile ($R_{wp}$) values, which quantifies the overall agreement between the experimental data ($I_i ^{obs}$) and the calculated intensities from the refined crystal structure ($I_i ^{calc}$) while accounting for the experimental uncertainties. This figure of merit is given by 
\begin{equation}
R_{wp}=\bigg(\sum_i ^n\frac{w_i (I_i ^{obs} - I_i ^{calc})^2}{\sum_i ^n w_i (I_i ^obs)^2}\bigg)^{1/2} \times 100\%   
\end{equation}
where $w_i$ is the weighting factor that accounts for the experimental uncertainties. The different figures of merit implemented in GSAS II have been described in detail by Toby et.al. \cite{toby_gsas-ii_2013, rietveld_rietveld_2014, runcevski_rietveld_2021} 

\clearpage
\section{Rietveld refinements using the pseudo-orthorhombic $T_d^*$ phase}

Multiphase RR revealed that the orthorhombic phase with phase fraction $\sim$0.25 was present in our sample even at room temperature. Orthorhombic $T_d$ MoTe$_2$ is reported to be a Weyl semimetal with many non-trivial quantum properties, and it is unlikely that disorders in the sample result in such a phase. We explored the possibility of a pseudo-orthorhombic phase using $T_d^*$ as proposed in Ref. \cite{tao_appearance_2019}. The refined lattice parameters and atomic positions of the $T_d^*$ phase are in TABLE \ref{tab:refinedTdstar}. Multiphase RR with this phase yielded a comparable phase fraction and a better-fit quality than the one using the $T_d$ phase.

\begin{table}[!h]
\caption{\label{tab:refinedTdstar}Refined atomic positions in the pseudo-orthorhombic $T_d^*$ phase (S.G. P$2_1$/$m$). The refined lattice parameters were $a$ = 6.3164 \AA \ , $b$ = 3.4718 \AA \ , $c$ = 27.6022 \AA \ , and $\beta$ = 91.362$^{\circ}$. The goodness of fit for the multiphase refinement was $\chi^2$ = 5.56 and $R_w$ = 0.095.}
\begin{ruledtabular}
\begin{tabular}{cccccccc}
Atom& x & y & z & Atom & x & y & z\\
\hline
Mo1 & 0.7040 (4) & 0.7500 & 0.9962 (9) & Mo2 (4) & 0.8170 & 0.2500 & 0.2534 (9) \\
Mo3 & 0.3050 (4) & 0.2500 & 0.5062 (9) & Mo4 (5) & 0.3050 & 0.7500 & 0.2249 (12) \\
Tel & 0.0661 (29) & 0.7500 & 0.0558 (8) & Te2 (4) & 0.4760 & 0.2500 & 0.3035 (9) \\
Te3 & 0.5389 (31) & 0.2500 & 0.0604 (7) & Te4 (4) & -0.0110 & 0.7500 & 0.3098 (7) \\
Te5 & 0.1070 (3) & 0.2500 & 0.1823 (10) & Te6 (3) & 0.3290 & 0.7500 & 0.4337 (7) \\
Te7 & 0.6140 (3) & 0.7500 & 0.1945 (7) & Te8 (3) & 0.8710 & 0.2500 & 0.4465 (7) \\
\end{tabular}
\end{ruledtabular}
\end{table}

\begin{figure}[ht]
    \centering
    \includegraphics{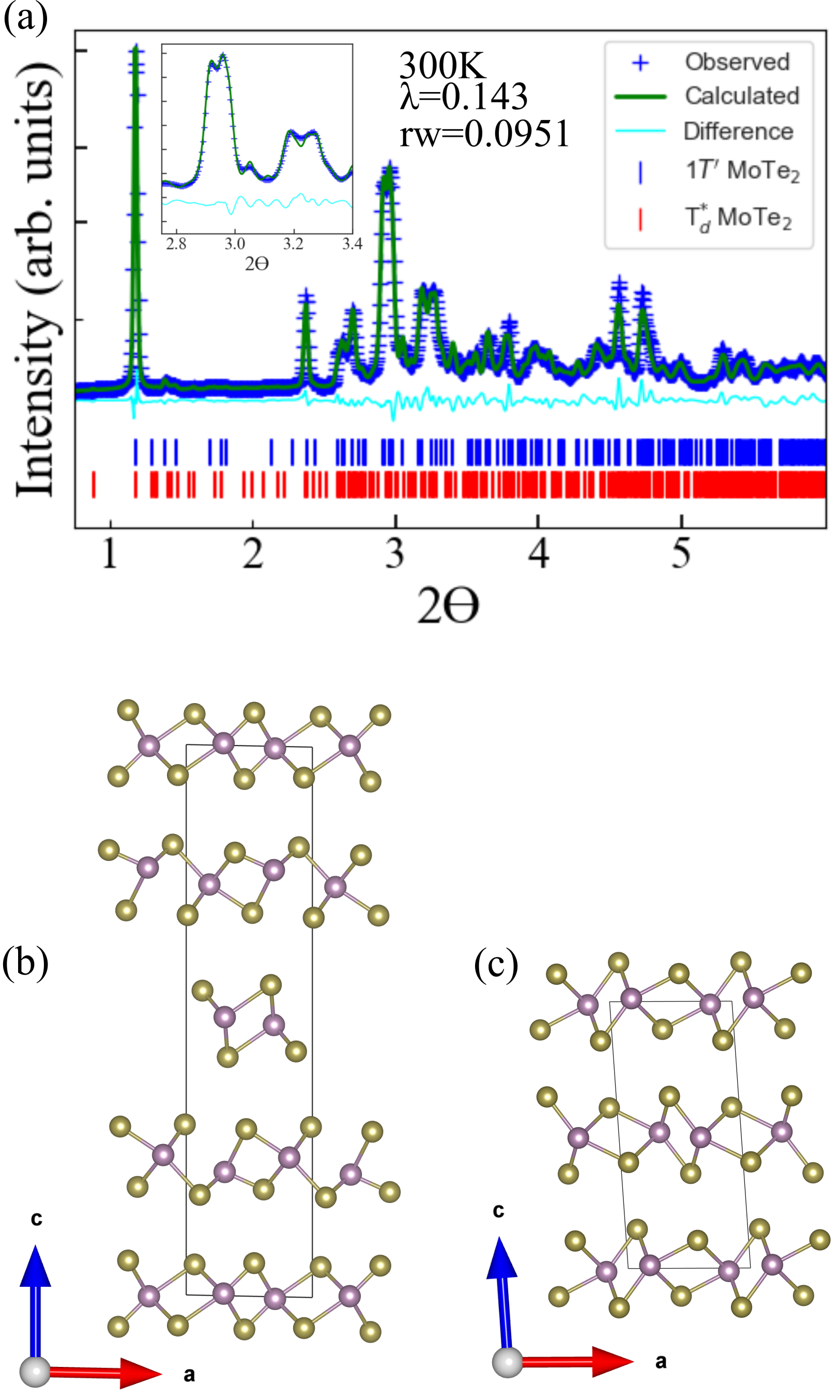}
    \label{TdStar}
    \caption{(a) Multiphase Rietveld refinement results using $T_d^*$ instead of the topological $T_d$ phase. The fit quality ($R_w$=0.095) is slightly better than the one using the $T_d$ phase ($R_w$=0.118). (b) Refined crystal structure for the pseudo-orthorhombic $T_d^*$ phase and (c) for the monoclinic $T_d$ phase after multiphase rietveld refinement.}
\end{figure}

\clearpage
\section{PDF Modeling}
The SPT from 1$T'$ $\rightarrow$ $T_d$ is suppressed over local scales, most likely due to the presence of stacking faults and the strong atom-atom correlation as revealed by RR of XRD data and small box modeling of PDF data. When using the orthorhombic $T_d$ phase for small box modeling of the PDF data, the fitting was not improved, and the $r_w$ value was greater than using the monoclinic 1$T'$ phase at all temperatures.

\begin{figure}[hb]
    \includegraphics{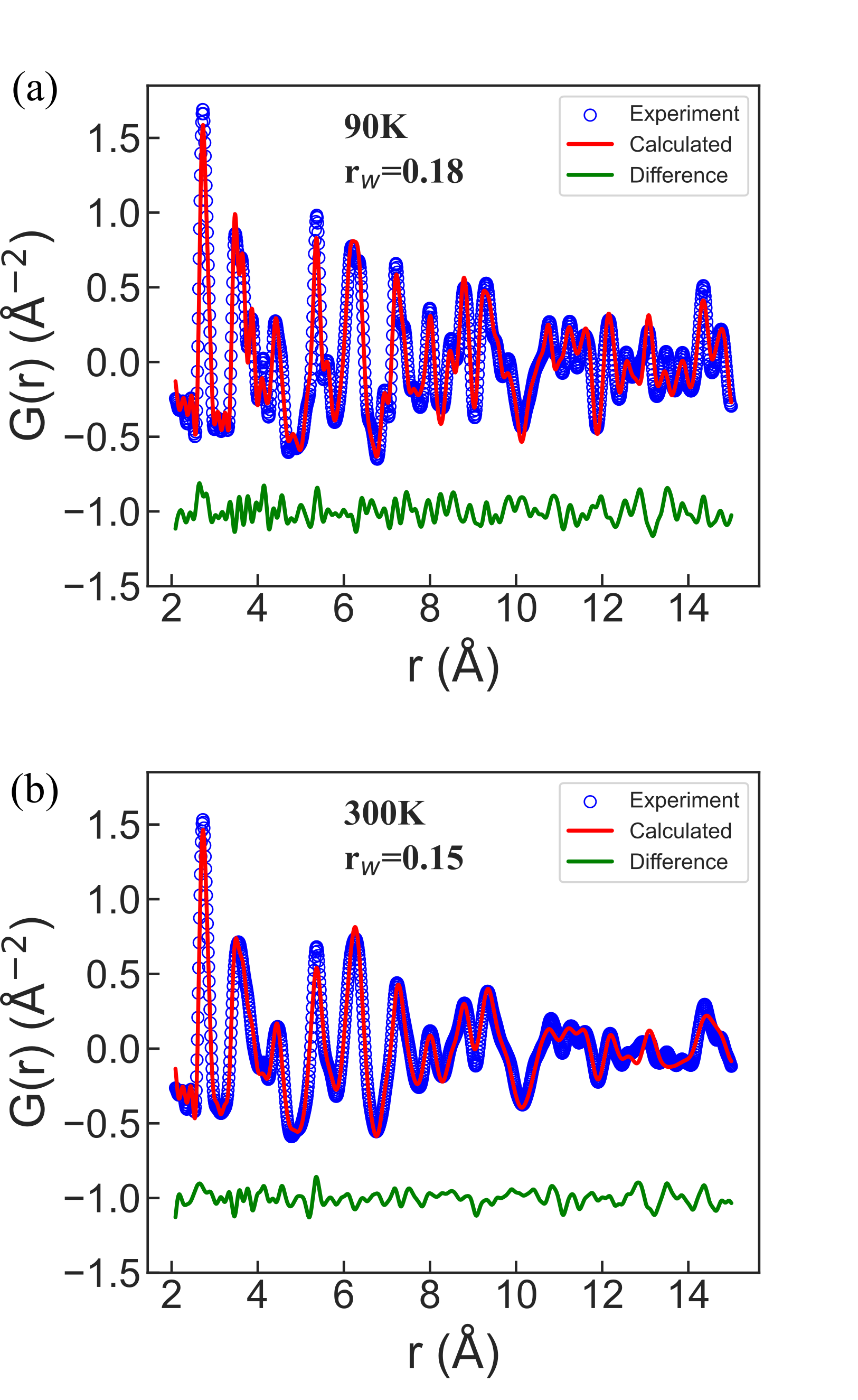}
    \caption{PDF Fits at a) 90 K and b) 300 K using the orthorhombic $T_d$ phase. The experimental data is represented by blue circles, simulated data by red curve, and the difference between them by the green curve. The difference curve is shifted along the -ve y-axis in the graph for clarity.} 
    \label{fig:TdFits}
\end{figure}

\clearpage
The ADPs obtained from the Rietveld-like refinements for the PDF data are represented by the actual displacement of atoms for large box modeling (Reverse Monte Carlo fitting or RMC refinement). For visualization, the supercell obtained after RMC refinement is collapsed into the unit cell to obtain an atomic point-cloud distribution. 

\begin{figure}[ht]
    \centering
     \includegraphics[scale=0.8]{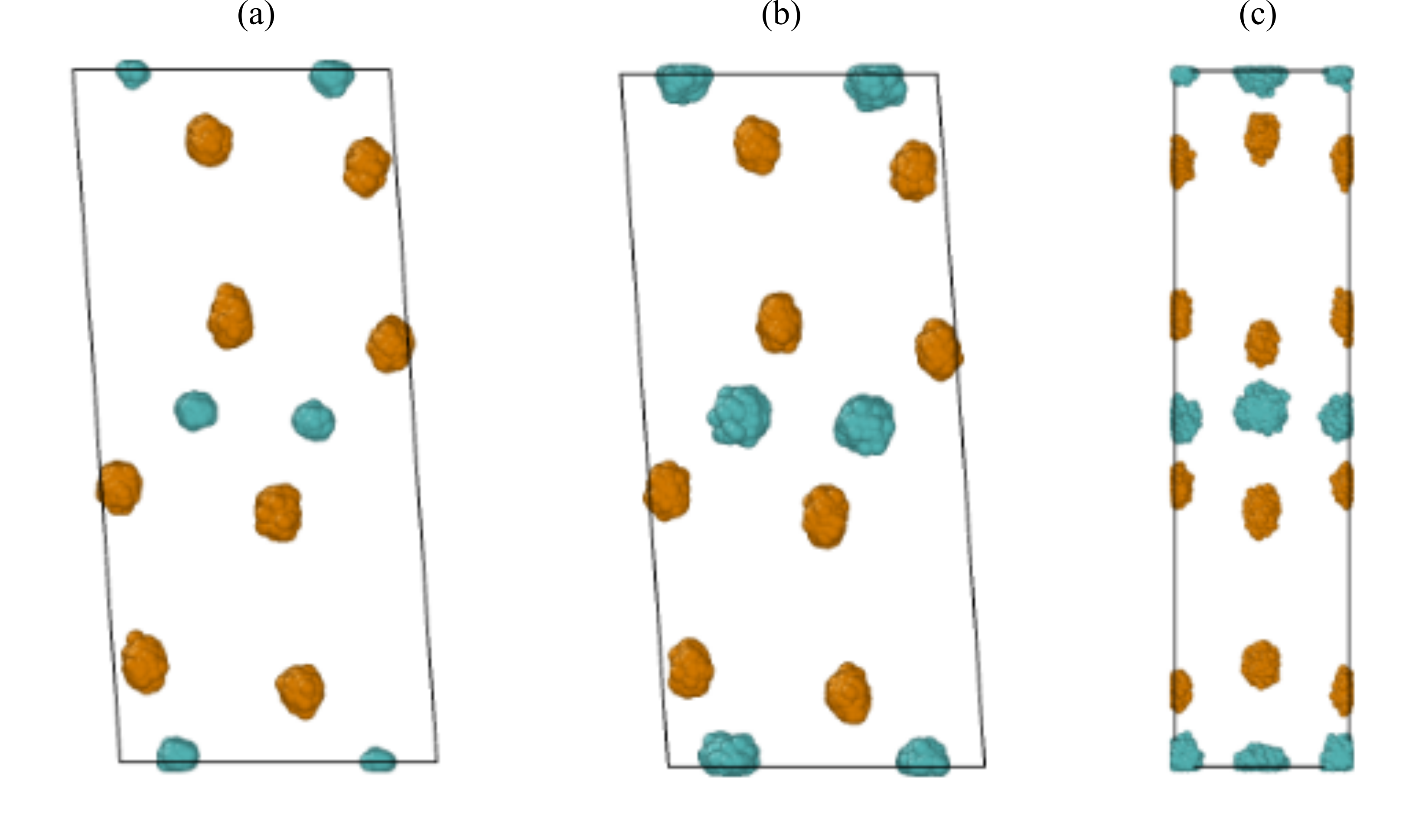}
     \caption{Projection of the atomic positions of MoTe$_2$ from the RMC supercell into the 1$T$' original unit cell for the local structure at (a) 90 K, (b) 300 K, and (c) T$_d$ unit cell at 300K. Atoms are colored as follows: Mo: cyan, Te: golden}
     \label{collapsedCell}
\end{figure}

\clearpage
\section{Layering Operations}

Different layered phases of MoTe$_{2}$, including the 1$T'$ and the $T_{d}$ phases, can be visualized as crystal structures by applying different operations to the layering order. In the simplest scenario, the stacking starts with a layer, and the subsequent layers along the positive $z$-direction are obtained by applying a certain set of operations. The difference between the two phases now would be the set of operations applied to it. The set of stacking operations (described as $A$ and $B$) that results in different topological phases of MoTe$_2$ is discussed in detail in Ref. \cite{schneeloch_emergence_2019}.
\begin{figure}[hb]
    \centering
    \includegraphics[scale=0.85]{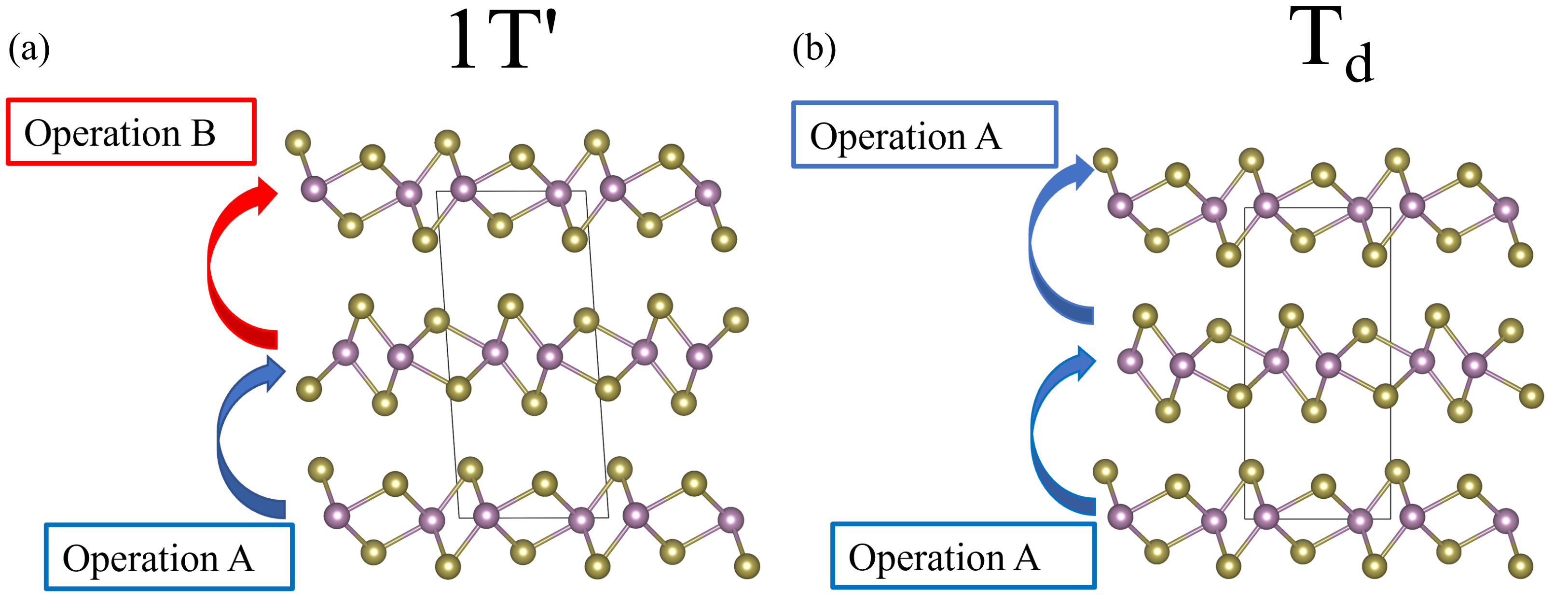}
    \caption{Different stacking (layering) arrangements can produce two different topological phases of MoTe$_{2}$. (a) AB stacking results in 1$T'$ phase while (b) AA stacking results in $T_d$ phase.}
    \label{fig:layeringOps}
\end{figure}

\clearpage
\bibliographystyle{unsrt} 
\bibliography{supplement}